# The GALAH Survey: Neutron-Capture Elemental Abundances for 350 000 *Gaia*-RVS spectra and the Chemodynamics of Accreted Structures

Pradosh Barun Das,[1,2,3★] Daniel B. Zucker,[1,2] Aldo Mura-Guzmán,[1,2] Nicholas W. Borsato,[1,2,4] Gayandhi M. De Silva,[1,2,5] Sven Buder,[5,6] Diane Feuillet,[7,8] Thomas Nordlander,[9] Melissa K. Ness,[5] Sarah L. Martell,[10] Janez Kos,[11] Joss Bland-Hawthorn,[12] Ken C. Freeman,[5] Andrew R. Casey,[13,14] Geraint F. Lewis,[12] Dennis Stello,[10,12] Richard de Grijs[1,2,15] and the GALAH Collaboration†

*Affiliations are listed at the end of the paper*

Accepted 2026 May 29. Received 2026 May 27; in original form 2026 February 2

## ABSTRACT

We present a comprehensive data-driven spectroscopic analysis of 357 415 red giant stars using *Gaia* DR3 Radial Velocity Spectrometer (RVS) spectra (8460–8700 Å; $R \approx 11\,500$), aimed at deriving homogenous stellar parameters and elemental abundances (collectively referred to as stellar labels). We employ *The Cannon*, a generative model based on 2747 giants in common with GALactic Archaeology with HERMES (GALAH) DR4, adopting GALAH labels ($R \approx 28\,000$) for training. The resulting model predicts 11 stellar labels for RVS giants: effective temperature ($T_{\rm eff}$), surface gravity ($\log g$), projected rotational velocity ($v \sin i$), and abundances of [Fe/H], [Ca/Fe], [Si/Fe], [Ni/Fe], [Ti/Fe], as well as the neutron-capture elements [Zr/Fe], [Ce/Fe], and [Nd/Fe]. Building on these results, we develop a probabilistic framework to chemically identify debris from the *Gaia*–Sausage–Enceladus (GSE) accretion event. A logistic regression classifier, optimized via Markov chain Monte Carlo sampling and trained on a small reference sample of GSE members and comparison stars, identifies stars with high GSE membership probabilities based solely on their chemical abundances, with the resulting candidates exhibiting distinctive abundance-ratio patterns, including [Ca/Ti], [Ti/Ce], and [Nd/Zr]. Applying independent kinematic constraints yields a robust sample of GSE candidates, demonstrating that the characteristic chemical signatures remain consistent after applying these constraints. This work demonstrates the potential of data-driven analysis techniques to extract detailed chemical information from medium-resolution spectra and establishes a framework for tracing Galactic accretion events using chemical abundances.

## 1 INTRODUCTION

Deciphering the formation and chemical evolution history of the Milky Way requires precise measurements of stellar parameters and elemental abundances across diverse Galactic populations. Large-scale spectroscopic surveys such as GALAH (GALactic Archaeology with HERMES; G. M. De Silva et al. 2015; S. L. Martell et al. 2017; S. Buder et al. 2018, 2021, 2025), APOGEE (Apache Point Observatory Galactic Evolution Experiment; S. R. Majewski et al. 2017), and *Gaia*-ESO (G. Gilmore et al. 2012; S. Randich, G. Gilmore & Gaia-ESO Consortium 2013) have provided transformative insights into the chemodynamical structure of the Galaxy. Complementing these, the European Space Agency (ESA) *Gaia* mission (Gaia Collaboration 2016, 2018, 2023a) has yielded astrometric and photometric data for over a billion stars, with radial velocities and medium-resolution spectra from the Radial Velocity Spectrometer (RVS; M. Cropper et al. 2018; A. Recio-Blanco et al. 2016, 2023) for ∼33 million stars, enabling chemodynamical studies of the Milky Way (Gaia Collaboration 2023b). However, the limited spectral resolution and relatively narrow wavelength range of *Gaia*-RVS ($R \approx 11\,500$; 8460–8700 Å) present significant challenges for detailed chemical abundance analyses using traditional methods, such as equivalent width measurements or spectral synthesis fitting, which require either higher spectral resolution or a wider wavelength baseline to resolve individual spectral features and constrain stellar parameters independently.

Data-driven approaches, on the other hand, can overcome these limitations to extract reliable stellar labels from lower-resolution spectra by exploiting the full stellar information encoded in each spectrum. One such method is *The Cannon* (M. Ness et al. 2015; A. R. Casey et al. 2016; A. Y. Q. Ho et al. 2017a), a generative model that uses a well-characterized training set to model the normalized spectral flux as a function of stellar labels, and infer those stellar labels for a much larger set. This approach has proven effective in combining the precision of high-resolution data sets with the scale of more extensive but noisier spectroscopic catalogues.

★ E-mail: pradoshbarun.das@mq.edu.au, pbdrohan@gmail.com
† https://www.galah-survey.org

Since its initial application to APOGEE DR10 using a training sample of only 542 reference stars (M. Ness et al. 2015), *The Cannon* has been extensively validated across a wide range of spectroscopic surveys. A. R. Casey et al. (2016) extended *The Cannon* to detailed chemical abundance analyses using APOGEE spectra, delivering 17 stellar labels (including 15 elemental abundances) for 87 563 red giant stars. A. Y. Q. Ho et al. (2017a) applied *The Cannon* to transfer APOGEE labels ($R\approx$ 22 500) to 454 180 giants in the LAMOST (Large Sky Area Multi-Object Fibre Spectroscopic Telescope) survey ($R\approx$ 1800; A.-L. Luo et al. 2015) using 9952 common stars as training data, eliminating systematic label inconsistencies between surveys and producing [$\alpha$/M] measurements from LAMOST spectra. A. Y. Q. Ho et al. (2017b) used *The Cannon* to derive stellar masses and ages for 230 000 LAMOST giant stars from their carbon and nitrogen abundance signatures. S. Buder et al. (2018) applied *The Cannon* to 342 682 GALAH stars to measure 23 elemental abundances, demonstrating robust label transfer within a high-resolution optical spectroscopic survey. A. Wheeler et al. (2020) estimated abundances for nearly 3.9 million LAMOST stellar spectra across five nucleosynthetic channels using GALAH as reference data, whilst G. Nandakumar et al. (2022) employed *The Cannon* to construct a combined APOGEE–GALAH catalogue, demonstrating its utility to homogenize stellar labels across independent surveys. More recently, P. B. Das et al. (in preparation) applied *The Cannon* to $\sim$ 36 000 *Gaia*-ESO survey HR15N spectra ($R\approx$ 19 200; $\lambda\approx$ 6440–6815 Å) to derive Ba and Eu abundances, demonstrating the capability of the method to recover neutron-capture elemental abundances from intermediate-resolution spectra.

*The Cannon* is also well suited to *Gaia* DR3 RVS spectra, which, despite their lower spectral resolution, encompass key diagnostic features, including the near-infrared Ca II triplet and multiple absorption lines from $\alpha$-, Fe-peak, and neutron-capture elements. P. B. Das et al. (2025) applied *The Cannon* to derive stellar parameters and elemental abundances for approximately 800 000 *Gaia*-RVS spectra using GALAH DR4 ($R\approx$ 28 000) as the reference set, trained on 14 484 stars in common between the two surveys. The resulting abundance distributions, particularly the [Ti/Fe]–[Fe/H] plane, clearly recovered the characteristic $\alpha$-element bimodality of Milky Way disc stars. Validation using open and globular cluster stars demonstrated metallicity precisions of approximately 0.02–0.10 dex, confirming that *The Cannon* can derive reliable stellar abundances from medium-resolution RVS spectra with high precision (see fig. 10 in P. B. Das et al. 2025). These results are consistent with independent analyses of the same data set, such as G. Guiglion et al. (2024), who employed a Convolutional Neural Network (CNN) to derive stellar parameters, metallicities, and [$\alpha$/Fe] ratios from *Gaia* DR3 RVS spectra, also recovering the $\alpha$-bimodality of the Galactic disc.

In this work, we apply *The Cannon* to a curated sample of red giant stars with *Gaia* DR3 RVS spectra, using stellar parameters and elemental abundances from GALAH DR4 as the training labels. In particular, we extend the label space to include the neutron-capture elements Zr, Ce, and Nd – elements whose features are weak and often blended in medium-resolution spectra but which can provide valuable constraints on the chemical evolution of the Galaxy (C. Sneden, J. J. Cowan & R. Gallino 2008; A. Recio-Blanco et al. 2023). These elements are common tracers of neutron-capture enrichment: Zr and Ce are dominated by the *s*-process at near-solar metallicities, while Nd receives significant contributions from both *s*- and *r*-processes (e.g. C. Sneden et al. 2008). Their abundance patterns are therefore especially useful for disentangling *in situ* versus accreted stellar populations and for constraining the time-scales of chemical enrichment in different Galactic components. Our sample of 357 415 red giant stars drawn from the thin disc, thick disc, and halo enables a statistically robust investigation of Galactic chemical evolution and accreted stellar populations across diverse kinematic and chemical environments.

A key application of these chemical abundance labels is the identification of stars associated with the *Gaia*–Sausage–Enceladus (GSE) merger event (V. Belokurov et al. 2018; A. Helmi et al. 2018), one of the most significant accretion episodes in our Galaxy. This event deposited a population of stars with low angular momentum and highly eccentric orbits into the stellar halo. While previous GSE membership identification has relied primarily on kinematics, recent studies have emphasized the value of chemical abundance patterns for tracing accreted populations. Early work by P. E. Nissen & W. J. Schuster (2010) identified two chemically distinct halo populations in the solar neighbourhood via their [Mg/Fe], [Na/Fe], and [Ni/Fe] ratios. Subsequent work extended these diagnostics: K. Hawkins et al. (2015), P. Das, K. Hawkins & P. Jofré (2020), and A. Carrillo et al. (2022) used [Al/Fe] versus [Mg/Mn] to separate accreted stars, while S. Buder et al. (2022) demonstrated the utility of [Na/Fe] versus [Mg/Mn] in identifying GSE members. However, in this study, we do not have measurements for all of these key elements. Instead, to identify new potential GSE candidates, we build a probabilistic classification model based solely on available chemical abundances. Using logistic regression optimized via Markov chain Monte Carlo (MCMC), we infer GSE membership probabilities for our full sample. This is followed by a kinematic refinement based on orbital parameters derived from Galactocentric velocities and integrals of motion. By incorporating additional elemental abundances and a larger sample of stars, our study aims to better characterize the complex spatial, chemical, and dynamical structure of GSE, potentially reconciling differences in the metallicity distributions reported in previous studies (see e.g. fig. 10 in S. Buder et al. 2022).

The structure of this paper is as follows: Section 2 describes the dataset used in our *The Cannon* model, including the *Gaia* DR3 RVS spectra and GALAH DR4 labels. A description of *The Cannon*'s algorithm, including an overview of various selection criteria for the training sample as well as the flagging procedures of the red giant stars in the sample, is provided in Section 3. In Section 4, we assess the significance of *The Cannon*-predicted elemental abundances of neutron-capture elements in RVS spectra using differential analysis and cross-correlation techniques. Then, using the final *The Cannon* catalogue of stellar abundances, Section 5 discusses the empirical probabilistic model developed using MCMC methods to chemically identify

GSE candidates based solely on elemental abundances, followed by a dynamical analysis to refine the GSE candidate sample. Finally, Section 6 summarizes our findings and implications for upcoming spectroscopic data sets such as *Gaia* DR4.

## 2 DATA

The primary data set utilized in this study is the catalogue[1] from P. B. Das et al. (2025), which provides *The Cannon* predictions for eight stellar labels – three stellar parameters and five elemental abundances – for a total of 796 633 *Gaia*-RVS stars. We applied the `flag_cannon = 0` quality criteria to ensure reliable stellar parameters, and used the $\log g$ values from P. B. Das et al. (2025) to select our initial sample of 357 415 giant stars ($\log g < 3.5$ dex). In our analysis, we use the corresponding stellar spectra from *Gaia* DR3 RVS[2] and incorporate stellar parameters ($T_{\rm eff}$, $\log g$, $v \sin i$) along with chemical abundances ([Fe/H], [Ca/Fe], [Si/Fe], [Ni/Fe], [Ti/Fe], [Ce/Fe], [Zr/Fe], and [Nd/Fe]) from GALAH DR4 (S. Buder et al. 2025) to train the model using *The Cannon*.

### 2.1 *Gaia*-RVS

The *Gaia*-RVS is a near-infrared spectrograph (M. Cropper et al. 2018; Gaia Collaboration 2018, 2023a) on the ESA *Gaia* satellite, designed to obtain radial velocities, spectral classifications, and astrophysical parameters for bright stars as part of *Gaia*'s astrometric and photometric survey mission. Spanning 8460–8700 Å at a resolving power of $R \approx 11\,500$, RVS targets the Ca II triplet region, suitable for determining stellar kinematics and metallicities across a wide range of spectral types.

RVS spectra are acquired for stars brighter than $G \sim 16$ mag, with radial velocities for progressively fainter and more complex targets to be published in successive data releases. *Gaia* DR3 provides radial velocities for over 33 million stars (A. Recio-Blanco et al. 2023), along with median RVS spectra for $\sim 1$ million bright sources. These spectra are corrected for background and instrumental effects and co-added from multiple *Gaia* transits.

In addition to radial velocities, the RVS spectra also yield stellar atmospheric parameters and chemical abundances through the *Gaia* astrophysical parameter inference pipeline, GSP-Spec (A. Recio-Blanco et al. 2016, 2023). For bright stars ($G \lesssim 12$ mag), *Gaia* DR3 provides estimates of quantities such as $T_{\rm eff}$, $\log g$, [M/H], and abundances for a limited number of elements based on comparisons with synthetic spectral grids (A. Recio-Blanco et al. 2023). The combination of RVS data with *Gaia* astrometry and photometry enables full six-dimensional phase-space mapping and chemodynamical studies of the Galaxy, particularly in the Solar neighbourhood.

### 2.2 GALAH Data Release 4

The GALAH survey is a high-resolution spectroscopic initiative conducted using the HERMES spectrograph (High Efficiency and Resolution Multi-Element Spectrograph; J. Brzeski, S. Case & L. Gers 2011; J. Heijmans et al. 2012; A. Sheinis et al. 2015) on the 3.9-m Anglo-Australian Telescope, operating at $R \approx 28\,000$. HERMES collects spectra in four discrete optical channels – 4713–4903 Å (blue), 5648–5873 Å (green), 6478–6737 Å (red), and 7585–7887 Å (near-infrared) – optimized for elemental abundance analysis. The instrument is fibre-fed by the 2-degree Field (2dF) positioner (I. J. Lewis et al. 2002), allowing simultaneous observation of up to 392 stars over a 2° field of view. GALAH comprises two observational phases with complementary targeting strategies (S. Buder et al. 2025): Phase-1 (DR1–DR3) targets nearby stars predominantly within ∼4 kpc of the Sun whilst avoiding the Galactic plane ($|b| > 10°$), whereas Phase-2 (incorporated into DR4) explicitly targets the Galactic plane at low Galactic latitudes to constrain the structure and chemical signatures of inner-disc and bulge populations. GALAH DR4 additionally includes dedicated observations of more than 300 stellar clusters, providing a diverse sample spanning from the local solar neighbourhood to the inner Galaxy.

GALAH DR4 (S. Buder et al. 2025) provides spectra, stellar parameters, and elemental abundances for 917 588 stars. It incorporates data from GALAH Phases 1 and 2, *TESS*–HERMES (S. Sharma et al. 2018), *K2*–HERMES (S. Sharma et al. 2019), and observations of open and globular clusters. The catalogue includes abundances for up to 32 elements for these stars. For this study, we select a sub-set of GALAH DR4 stars that meet quality control criteria: `flag_sp`= 0 (valid spectral synthesis solution with no spectral peculiarities), `flag_fe_h`= 0 and `flag_x_fe`= 0, where *x* refers to each chemical element of interest. This ensures reliable stellar labels for training.

## 3 METHOD

### 3.1 *The Cannon*

*The Cannon* (M. Ness et al. 2015; A. R. Casey et al. 2016) is a probabilistic, data-driven generative model designed to estimate stellar atmospheric parameters and elemental abundances by modelling normalized stellar spectra as a function of stellar labels. It relies on a reference set of spectra for which high-fidelity labels are known. Using these, it trains a forward model that predicts the flux at each pixel as a smooth function of the stellar labels. In its original formulation, the model is trained by optimising the likelihood of the observed flux at each wavelength pixel, assuming that the flux at wavelength $\lambda$ for star $n$ can be expressed as

$$G_{n\lambda} = f(l_n) \cdot \theta_\lambda + \epsilon_{n\lambda}. \tag{1}$$

Here, $f(l_n)$ is a vectorizing function of the stellar labels $l_n$ (typically a quadratic polynomial), $\theta_\lambda$ is the coefficient vector at pixel $\lambda$, and $\epsilon_{n\lambda}$ is the residual. During training, $\theta_\lambda$ and an additional intrinsic scatter term $s_\lambda$ are optimized to maximize the total log-likelihood across the training set. The residual $\epsilon_{n\lambda}$ accounts for both the observational uncertainty $\sigma_{n\lambda}$, which represents the per-pixel flux error, and the intrinsic model scatter $s_\lambda$, which captures astrophysical variance not explained by the model. These contributions combine in quadrature to give the total variance

$$\sigma^2_{{\rm tot},n\lambda} = \sigma^2_{n\lambda} + s^2_\lambda. \tag{2}$$

Under the Gaussian noise assumption, the residual $\epsilon_{n\lambda}$ is therefore distributed as

$$\epsilon_{n\lambda} \sim \mathcal{N}(0, \sigma^2_{{\rm tot},n\lambda}). \tag{3}$$

In other words, the observed residuals reflect both measurement noise and intrinsic scatter in the model. The training procedure involves maximizing the following log-likelihood for each

[1] https://cdsarc.cds.unistra.fr/viz-bin/cat/J/MNRAS/538/605
[2] https://doi.org/10.17876/gaia/dr.3/54

wavelength independently

$$\ln p(G_{n\lambda}|\theta_\lambda, f(l_n), s_\lambda^2) = -\frac{[G_{n\lambda} - f(l_n)\cdot\theta_\lambda]^2}{2(\sigma_{n\lambda}^2 + s_\lambda^2)} - \frac{1}{2}\ln(\sigma_{n\lambda}^2 + s_\lambda^2). \quad (4)$$

Once trained, the coefficients $\theta_\lambda$ and $s_\lambda$ are fixed. For an unlabelled test spectrum, stellar labels are inferred by maximising the likelihood of the observed flux with respect to the labels

$$l_m \leftarrow \arg\min_{l} \sum_{\lambda=1}^{N_\lambda} \frac{[G_{m\lambda} - f(l)\cdot\theta_\lambda]^2}{\sigma_{m\lambda}^2 + s_\lambda^2}. \quad (5)$$

This formulation allows *The Cannon* to robustly transfer label information from higher signal-to-noise ratio (SNR) training data to lower SNR test spectra without requiring the training and test sets to be statistically identical. Moreover, its fully generative design makes it amenable to cross-calibration across heterogeneous spectroscopic surveys (e.g. APOGEE, LAMOST, GALAH, *Gaia*-DR3 RVS), as demonstrated in applications such as A. Y. Q. Ho et al. (2017a), S. Buder et al. (2018), A. Wheeler et al. (2020), G. Nandakumar et al. (2022), and P. B. Das et al. (2025). In this work, we adopt the framework of *The Cannon* [3] (M. Ness et al. 2015; A. R. Casey et al. 2016), constructing a second-order polynomial model in label space and training it on a selected reference set. This enables us to propagate labels to our target sample while maintaining physical interpretability throughout the sample.

### 3.2 Training data set

From the total sample of 357 415 giant stars ($\log g < 3.5$ dex, see Section 2) selected from the catalogue (P. B. Das et al. 2025), we further refined the selection conditions to create a training sample focused on extracting more localized chemical information and individual elemental abundances. The training data set was generated by adhering to the quality criteria explained below and additional [Fe/H] quality cuts for the training sample.

To ensure a representative training sample, including stars over a broad metallicity range, a comparable number of stars were selected in different metallicity bins using multiple criteria. Stars from the *Gaia*-RVS data set with GALAH metallicity labels of $[\mathrm{Fe/H}] \leq -0.8$ dex were required to have a *Gaia*-RVS spectral SNR ≥25 and GALAH spectral SNR (`snr_px_ccd3`) ≥ 30. Additional selection criteria included GALAH effective temperature ($T_{\mathrm{eff}}$) < 7000 K, *Gaia* renormalized unit weight errors (`ruwe`) < 1.2, GALAH chi-squared values (`chi2_sp`) < 4, positive *Gaia* parallaxes (`parallax`) > 0 mas, and satisfying flags from the GALAH data reduction process, which ensured reliability of both the spectral data and the derived elemental abundances. In addition to this, we also included stars with GALAH metallicity labels for $[\mathrm{Fe/H}] > -0.8$ dex with more stringent criteria in addition to the above-mentioned conditions: *Gaia*-RVS SNR ≥ 50, GALAH SNR ≥ 50 across all four CCD arms, and small uncertainties associated with GALAH DR4 chemical abundances (≤0.10 dex) and stellar parameters ($\mathrm{e_}T_{\mathrm{eff}} \leq 80$ K, and $\mathrm{e_}\log g \leq 0.10$ dex) to ensure data quality and reliability. Following these cuts, the new training sample comprised 2747 stars, spanning three broad metallicity ranges: 833 stars with $-2.0 < [\mathrm{Fe/H}]$ (dex) $\leq -0.8$, 1165 stars with $-0.8 < [\mathrm{Fe/H}]$ (dex) $\leq 0$, and 749 stars with $[\mathrm{Fe/H}] > 0$ dex. Such distribution of stars in the metallicity space provides a comprehensive coverage that is essential for robust predictions in subsequent test samples (see Fig. 1).

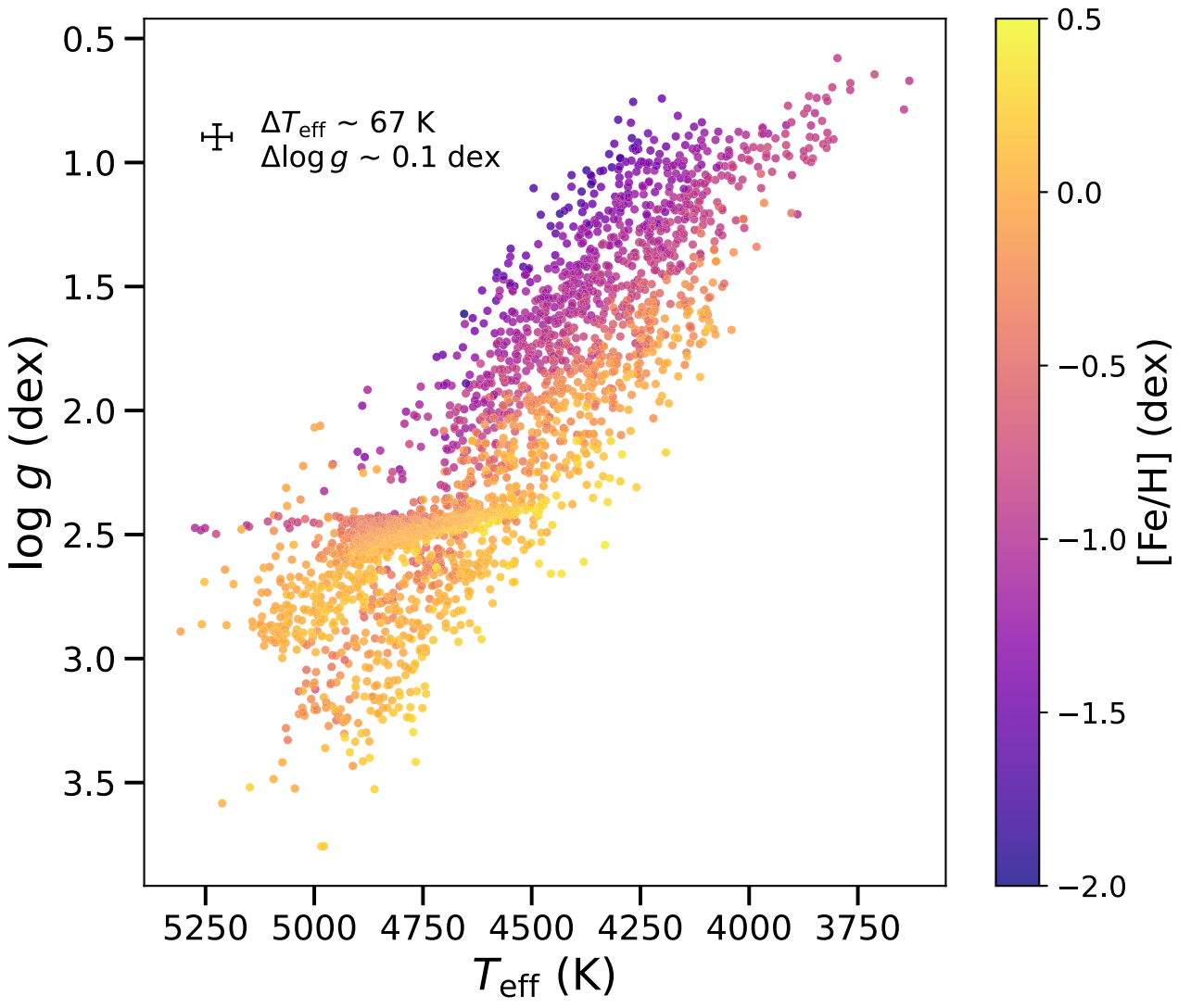

**Figure 1.** Kiel Diagram showing the GALAH values for $\log g$ versus $T_{\mathrm{eff}}$, colour-coded with [Fe/H] for the 2747 stars selected in the training sample. The representative error bar in the upper-left corner indicates the median uncertainties ($\Delta T_{\mathrm{eff}} \sim 67$ K and $\Delta \log g \sim 0.1$ dex) estimated in GALAH for the training sample.

For the analysis using *The Cannon*, a total of 11 stellar labels from GALAH DR4 were employed, including the stellar parameters ($T_{\mathrm{eff}}$, $\log g$, $v \sin i$) and elemental abundances ([Fe/H], [Ca/Fe], [Si/Fe], [Ni/Fe], and [Ti/Fe]), as well as the neutron-capture elements ([Zr/Fe], [Ce/Fe], and [Nd/Fe]). The trained model was then applied to predict these 11 stellar labels for the full sample of 357 415 giant stars ($\log g < 3.5$ dex). The choice of these elements was motivated by A. Recio-Blanco et al. (2023), who analysed these species within the *Gaia*-RVS wavelength range; we adopted the same set of elements following previous studies in this spectral region. Some elements present in the *Gaia*-RVS wavelength range were not included as training labels; Sulphur is not available in GALAH DR4, CN suffers from degeneracies between carbon and nitrogen abundances that prevent reliable measurements, and the Cr, Mg, Co, and O features in the RVS range are generally weak and provide limited diagnostic information.

Although the neutron-capture elements Zr, Ce, and Nd also exhibit weak lines, this study extends the application of *The Cannon* to the recovery of their abundances and evaluates its performance on medium-resolution, relatively low-SNR RVS spectra. To ensure that the stellar label space spanned by the training sample is representative of the test stars, we performed a 12-fold cross-validation using *The Cannon* to evaluate its performance in predicting stellar parameters and elemental abundances. For this analysis, the data set of 2747 stars was randomly divided into 12 sub-sets. In each iteration of the cross-validation, one sub-set was withheld as a validation set, while the model was trained on the remaining 11 sub-sets. The trained model was then employed to predict the stellar labels of the withheld sub-set. This methodology allowed us to calculate the root-mean-square error (RMSE) and biases, defined as the differences between the mean values predicted by *The Cannon* and the reference values from GALAH DR4, for the model trained on this data set. The RMSE and bias metrics for the stellar parameters derived from the training sample are presented in Table 1. These cross-validation statistics provide an estimate of the internal precision of the model within the training set.

[3] https://github.com/andycasey/AnniesLasso

**Table 1.** Biases and RMSE values of the 12-fold cross-validation test for the *Gaia*-RVS training sample consisting of 2747 giants ($\log g < 3.5$ dex) trained by *The Cannon* using stellar parameters and abundances from GALAH DR4 survey.

| Stellar label | RMSE | Bias |
|---|---|---|
| $T_{\rm eff}$ (K) | 70 | 12 |
| $\log g$ (dex) | 0.14 | 0.01 |
| $v \sin i$ (km s$^{-1}$) | 1.93 | 0.18 |
| [Fe/H] (dex) | 0.09 | 0.01 |
| [Ca/Fe] (dex) | 0.08 | 0.00 |
| [Si/Fe] (dex) | 0.06 | 0.01 |
| [Ni/Fe] (dex) | 0.04 | 0.00 |
| [Ti/Fe] (dex) | 0.05 | 0.00 |
| [Zr/Fe] (dex) | 0.15 | 0.02 |
| [Ce/Fe] (dex) | 0.18 | 0.02 |
| [Nd/Fe] (dex) | 0.15 | 0.01 |

Fig. 2 shows the one-to-one comparisons between the stellar labels predicted by *The Cannon* and the corresponding GALAH DR4 reference labels, while Fig. 3 presents the distributions of the residuals relative to the GALAH values. Overall, the predicted stellar labels exhibit good agreement with the reference labels, with small residual dispersions, demonstrating that the model is able to reproduce the GALAH DR4 labels with good precision. Although *The Cannon* provides robust label estimates across the full sample, predictions may be less reliable for spectra located near the boundaries of the training-label space or for lower SNR observations, where the associated uncertainties are correspondingly larger.

### 3.3 Application of *The Cannon* on *Gaia*-RVS spectra

Using the model developed with the training set, we applied *The Cannon* to predict stellar labels for the sample of 357 415 *Gaia*-RVS targets. The model was supplied with spectral fluxes and the associated flux errors for these *Gaia*-RVS targets, which allowed us to predict the stellar labels, including $T_{\rm eff}$, $\log g$, $v \sin i$, [Fe/H], [Ca/Fe], [Si/Fe], [Ni/Fe], [Ti/Fe], [Zr/Fe], [Ce/Fe], and [Nd/Fe] (hereafter, the labels). To ensure the validity of the model's predictions, it is essential to flag *The Cannon*'s estimates that lie outside the parameter range covered by the training set. This is essential for the accuracy of the predictions, which may decrease when extrapolating to regions outside the bounds of the training data. For each test star $n$ and each training star $j$, we define a normalized squared distance in label space (A. Y. Q. Ho et al. 2017a; S. Buder et al. 2018; G. Nandakumar et al. 2022; P. B. Das et al. 2025),

$$d_{nj}^2 = \sum_l \frac{\left(l_{n,l}^{\rm Cannon} - l_{j,l}^{\rm GALAH}\right)^2}{K_l^2}, \tag{6}$$

where the sum runs over all labels $l$ in the model, $l_{n,l}^{\rm Cannon}$ is *The Cannon*'s prediction for label $l$ of test star $n$, $l_{j,l}^{\rm GALAH}$ is the corresponding GALAH DR4 label for training star $j$, and $K_l$ is a characteristic scale for label $l$. We adopt $K_l$ equal to the RMSE values obtained for the labels from the 12-fold cross-validation of the training sample (Table 1), which provide an empirical measure of the typical label scatter in the training set.

For each test star $n$, we identify the set $\mathcal{N}_n$ of the $N_{\rm nn} = 10$ training stars with the smallest $d_{nj}$ values (i.e. the 10 nearest neighbours in label space). We then define the average nearest-neighbour distance

$$D_n = \frac{1}{N_{\rm nn}} \sum_{j \in \mathcal{N}_n} d_{nj}. \tag{7}$$

Small values of $D_n$ indicate that the test star lies within a well-sampled region of the training-set label space, while large values indicate potential extrapolation. Following P. B. Das et al. (2025), we adopt a global threshold $D_n < 2\sigma$ for all the labels together to define stars that are sufficiently close to the training-set manifold in this normalized label space. Stars satisfying $D_n < 2\sigma$ for all the labels are assigned `flag_cannon = 0`, while those with $D_n \geq 2\sigma$ are assigned `flag_cannon = 1`. Out of the 357 415 *Gaia*-RVS targets, 314 010 stars meet the $D_n < 2\sigma$ criterion and are considered to have labels within the limits of the total label space spanned by the training set. In addition to the global `flag_cannon`, we also define abundance-specific flags to assess the reliability of the inferred metallicity and individual elemental abundances. For [Fe/H], the corresponding nearest-neighbour distance is computed in the label space defined by $T_{\rm eff}$, $\log g$, $v \sin i$, and [Fe/H]. Stars with $D_n^{\rm (Fe)} < 2\sigma$ are assigned `flag_FeH = 0`, while those with $D_n^{\rm (Fe)} \geq 2\sigma$ are assigned `flag_FeH = 1`. For each element $X \in$ {Ca, Ti, Si, Ni, Nd, Ce, Zr}, we compute a corresponding nearest-neighbour distance $D_n^{(X)}$ in a reduced label space, including $T_{\rm eff}$, $\log g$, $v \sin i$, [Fe/H], and the specific abundance [$X$/Fe]. Stars with $D_n^{(X)} < 2\sigma$ are considered to lie within the well-sampled region of the training-set manifold for that specific element and are assigned `flag_XFe = 0`. Conversely, stars with $D_n^{(X)} \geq 2\sigma$ are flagged as `flag_XFe = 1`, indicating that the corresponding abundance may be affected by extrapolation beyond the training-set coverage.

We also note that the internal scatter provided by *The Cannon* (Section 3.1) describe the intrinsic model scatter in flux space and do not by themselves account for the full, SNR-dependent uncertainty of the inferred stellar labels. In addition to these internal contributions, there are external effects arising from finite SNR and other observational systematics, which can be parametrized as a function of spectral SNR (e.g. M. Ness et al. 2015; A. Wheeler et al. 2020; C. Manea et al. 2024). Following a similar approach to P. B. Das et al. (2025), we therefore generate $N_{\rm rep} = 10$ independent noise realizations of each *Gaia*-RVS spectrum in our test sample of 357 415 stars. For each realization, the flux at every wavelength pixel is redrawn from a Gaussian distribution with mean equal to the observed flux and standard deviation given by the reported per-pixel uncertainty. This procedure produces 10 synthetic spectra per star that incorporate SNR-dependent flux variations consistent with the observational noise. We apply *The Cannon* to each of these 10 noise realizations, obtaining 10 independent estimates of each of the 11 stellar labels for every star. For a given star $n$ and label $l$, we define the observational, SNR-dependent label uncertainty as the standard deviation of these noise-based estimates,

$$\sigma_{n,l}^{\rm obs} = {\rm StdDev}\left(l_{n,l}^{(1)}, l_{n,l}^{(2)}, \ldots, l_{n,l}^{(N_{\rm rep})}\right), \tag{8}$$

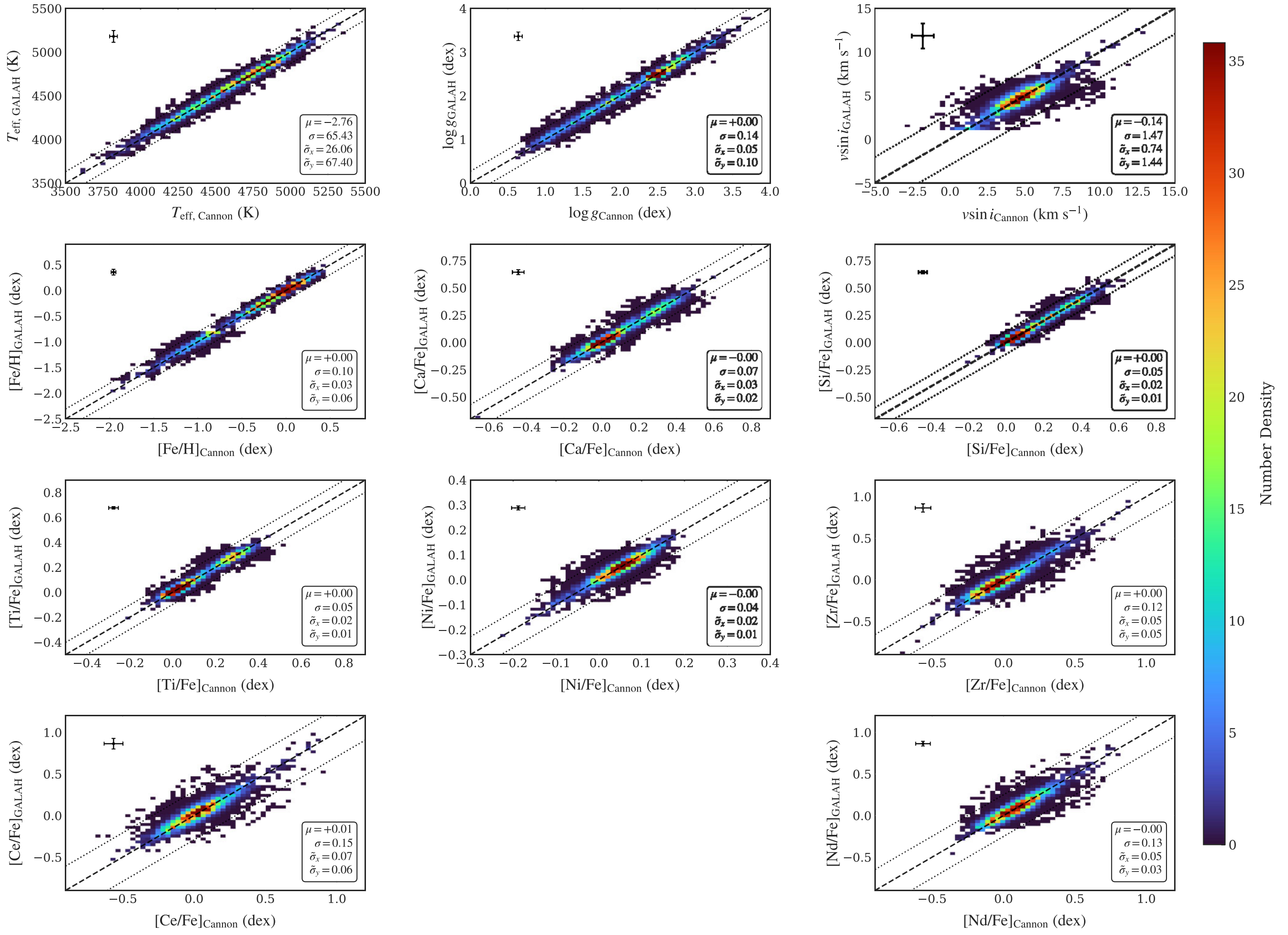

**Figure 2.** One-to-one relations for *The Cannon*-predicted stellar labels for the 2747 *Gaia*-RVS spectra of giants in the training sample. The horizontal axis indicates *The Cannon* predictions of the stellar labels, and the vertical axis shows the reference values from GALAH DR4. The black dashed line indicates the one-to-one relation, while the dotted lines mark $\pm 2\sigma$, where $\sigma$ is the standard deviation of the residuals (*The Cannon* predictions − GALAH DR4). The representative error bar in the upper left corner of each panel denotes the median uncertainties in the *The Cannon* predictions and the GALAH DR4 labels. $\mu$ and $\sigma$ indicate the mean and standard deviation of the residuals, respectively, while $\tilde{\sigma}_x$ and $\tilde{\sigma}_y$ represent the median uncertainties in the *The Cannon* and GALAH DR4 labels, respectively.

where $l_{n,l}^{(k)}$ denotes the value of label $l$ for star $n$ in the $k$th realization. In addition, *The Cannon* provides an internal covariance matrix in label space for each star, which reflects the curvature of the likelihood around the best-fitting solution. We characterize the corresponding internal uncertainty for label $l$ by the square root of the relevant diagonal element of this covariance matrix, denoted by $\sigma_{n,l}^{\rm int}$. The total variance in label $l$ for star $n$ is then obtained by adding the observational and internal contributions in quadrature,

$$\sigma_{n,l,\rm tot}^2 = \left(\sigma_{n,l}^{\rm obs}\right)^2 + \left(\sigma_{n,l}^{\rm int}\right)^2 . \tag{9}$$

The uncertainty reported in our catalogue for each stellar label is the square root of this total variance ($\sqrt{\sigma_{n,l,\rm tot}^2}$), which allows us to assess the reliability of the individual label measurements (see Table A1). We also compute reduced-$\chi^2$ values (`rchi_sq`) to assess the goodness-of-fit between *The Cannon*-predicted spectra and observed RVS spectra for each star (see Fig. A1). The `rchi_sq` distribution peaks at $\approx 1$ for the majority of stars, indicating that the model-predicted spectra are broadly similar to the observed spectra within the expected uncertainties.

The performance of the model across the full signal-to-noise range is assessed by examining the inferred abundance uncertainties as a function of spectral SNR for all stellar labels (see Fig. C1). We find a continuous increase in uncertainties toward lower SNR, consistent with the expected decrease in spectral information content. We do not observe any abrupt degradation at low SNR, suggesting that the model remains stable in this regime. This provides a quantitative estimate of the achievable precision as a function of SNR for all the stellar labels. This trend also offers a complementary guideline for using the catalogue. By relating the reported uncertainties to spectral SNR, the results provide an empirical basis for selecting SNR thresholds corresponding to a desired level of precision. Depending on the scientific application, more stringent SNR or uncertainty cuts may be adopted to construct high-precision samples, or more inclusive selections for statistical studies. The label uncertainty–SNR relation presented in Appendix C is intended to facilitate such choices.

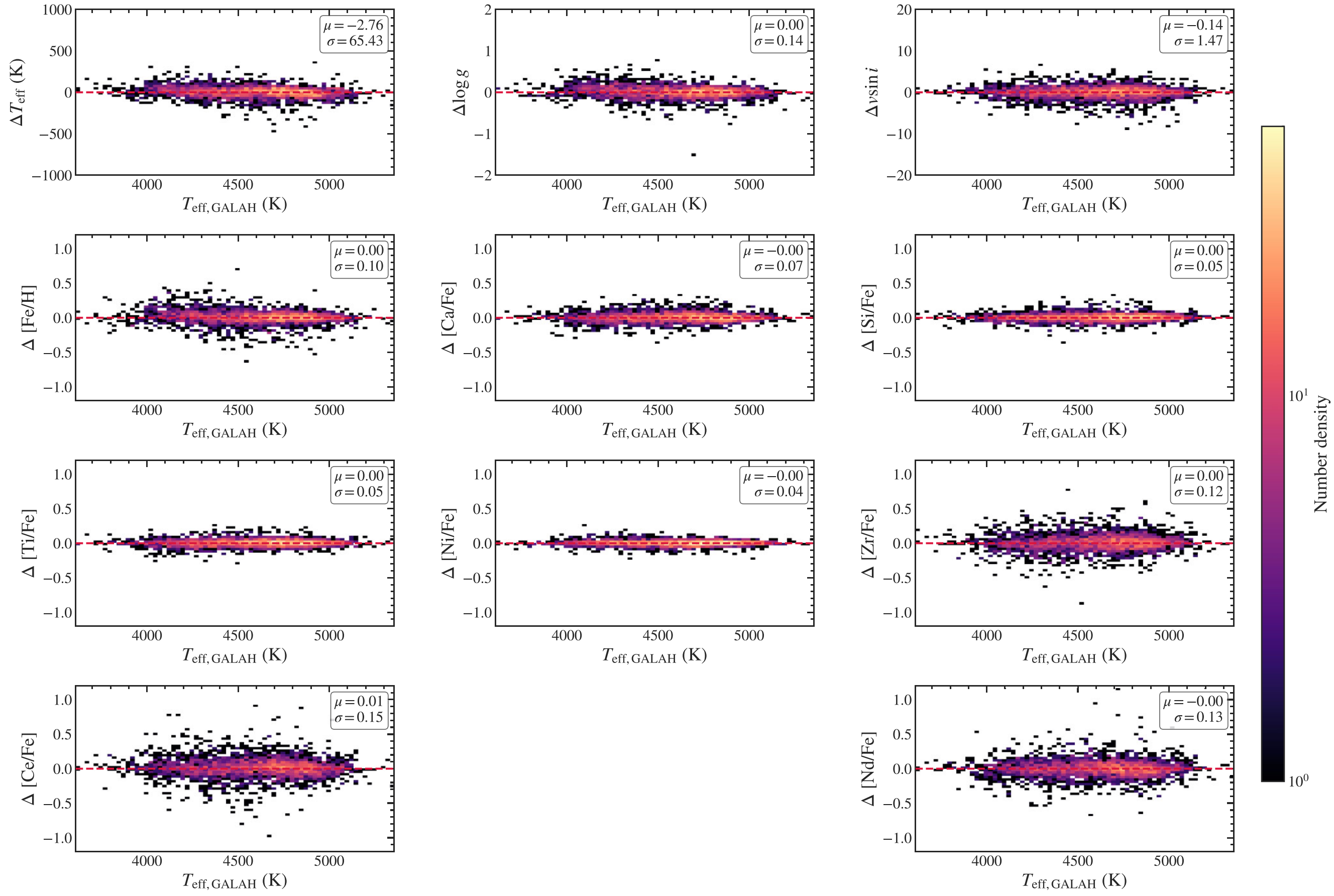

**Figure 3.** Residual values for *The Cannon*-predicted stellar labels (see Fig. 2). The horizontal axis shows $T_{\rm eff}$ values from GALAH DR4, and the vertical axis represents the difference between *The Cannon* predictions and GALAH DR4 values (*The Cannon*–GALAH DR4) for each of the stellar labels. $\mu$ and $\sigma$ represent the mean and standard deviation of the residual values, respectively.

## 4 ASSESSING NEUTRON-CAPTURE ABUNDANCES WITH The Cannon

The atomic transition lines of neutron capture elements like Ce, Zr, and Nd are typically quite weak compared to other absorption features at these wavelengths, such as other atomic transitions or molecular absorption. This difference in line strengths becomes a challenge when determining chemical abundances from medium-resolution data at low or moderate SNR. A common solution to this issue is to fit a synthetic spectrum or equivalent widths (EW) to the observed spectrum to determine its chemical composition. However, applying these methods to a large dataset like ours is unfeasible due to the weaknesses of the lines, and the computational resources required by these approaches. Data-driven methods such as *The Cannon* can address this issue by learning from a well-studied reference set of spectra – the training sample – for which a model is constructed using labels from another, typically higher resolution data set (in this case, GALAH DR4).

Given that this method does not explicitly measure the strengths of identified absorption features – or the many other steps entailed by more traditional analysis techniques – a key question arises: can *The Cannon* reliably recover chemical abundances from weak absorption lines, or could its outputs simply reflect trends in Galactic chemical evolution learned from the training sample? Specifically, in this case, can *The Cannon* actually measure the abundances of individual neutron-capture elements, or is it in fact simply tracing some composite of neutron-capture features, or even other elements whose abundances track with the neutron-capture elements? To address these questions, we inspected the *Gaia*-RVS spectral regions corresponding to these elements and evaluated the significance of the absorption features of interest.

### 4.1 Differential analysis of neutron-capture element absorption features

Within the *Gaia*-RVS wavelength range, Ce, Zr, and Nd have 8, 11, and 11 lines, respectively. The parameters of these *n*-capture lines are shown in Table D1. Note that for the species, we follow standard nomenclature to refer to the neutral states as X.0 and the ionized states as X.1, where X is the *Z*-number of the element.

Fig. 4 shows the observed spectrum from *Gaia*-RVS for a star in our sample with higher predicted Nd content from *The Cannon*, [Nd/Fe] = 0.61 dex (*Gaia* ID: 6809018744787777024, shown in blue). We compare this star with another of similar stellar parameters ($T_{\rm eff}$/ log $g$/[Fe/H]) but with a predicted Nd abundance of [Nd/Fe] = −0.02 dex (*Gaia* ID: 51670728611455360, in red). The flux errors reported by *Gaia*-RVS for these two stars are represented with shaded regions in their respective colours. In this figure, the top two panels display the locations of the Nd lines along the spectra, marked with dashed vertical lines and yellow shading for both stars (A, B, C, D...), followed below by the residuals of their difference. The bottom set of panels provides a closer look at these Nd lines and their residuals, with the line centre wavelengths indicated in the middle of each panel, again marked by a dashed vertical line and a yellow-shaded region.

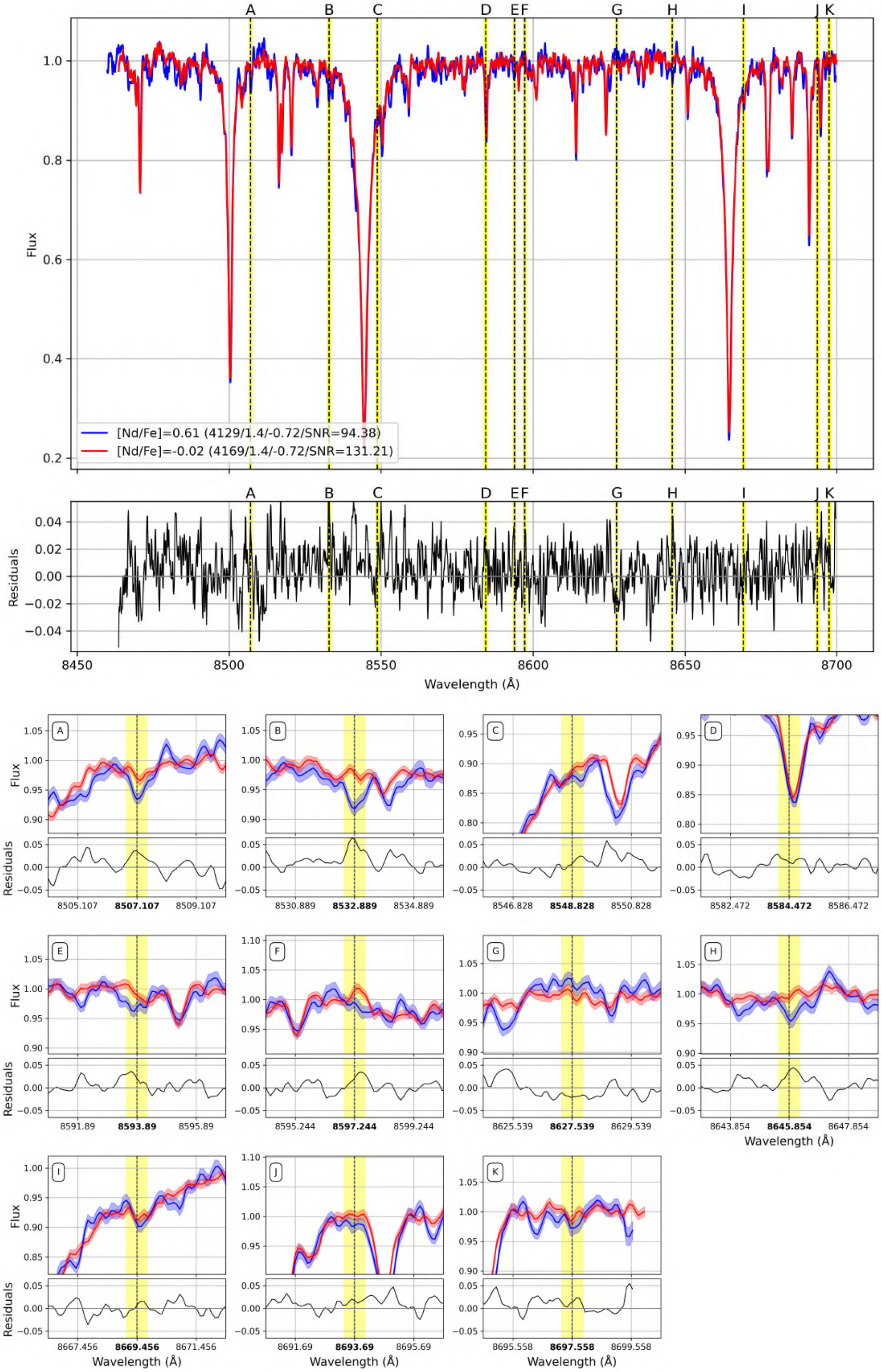

**Figure 4.** Comparison of two stars in our sample with similar stellar parameters and $\Delta$[Nd/Fe] $\sim$ 0.60 dex. The *Gaia*-RVS spectra correspond to a star with a higher Nd abundance predicted by *The Cannon*, [Nd/Fe] = 0.61 dex (blue; *Gaia* ID: 6809018744787777024) and another with [Nd/Fe] = −0.02 dex (red; *Gaia* ID: 5167072861145536 0). The Nd abundances and stellar parameters and SNR ($T_{\rm eff}$/ log $g$/[Fe/H]/SNR) of both stars are shown at the bottom left of the top-main panel. Shaded regions indicate flux uncertainties as reported by *Gaia*-RVS. The location of the Nd absorption features are marked by dashed vertical lines and yellow shading (A, B, C, D...). The middle panel shows flux residuals, while the bottom set of panels provide a zoom-in view of the Nd lines and their residuals with their line-centre wavelengths highlighted at the bottom.

This *pseudo*-differential analysis (for a more rigorous application of differential analysis of chemical abundances, see e.g. M. McKenzie et al. 2022; S. Monty et al. 2023) illustrates the flux differences between the two stars at $\Delta$[Nd/Fe] $\sim$ 0.60 dex. While small – on the level of at most a few per cent – these star-to-star flux differences at the line locations show that there is information for *The Cannon* to work with, even at the moderate resolution of *Gaia*-RVS.

### 4.2 Verifying the presence of neutron-capture element lines through cross-correlation

In certain cases, such as with ionized Calcium (Ca II), we can visually confirm the absorption lines, providing a clear verification of the model's accuracy. Additionally, cross-checking the abundance values with the GALAH DR4 test set further demonstrates that our model is a reasonable fit to the data. However, visually confirming the absorption lines for neutron-capture elements present in the *Gaia*-RVS spectra (such as Zr, Ce, and Nd) is challenging given the weakness of the lines (see e.g. Fig. 4) and the low SNR of the *Gaia*-RVS spectra.

To address this issue, we use the cross-correlation technique (I. A. G. Snellen et al. 2010). This method has previously been used to detect species in the noise-dominated absorption and emission spectra of exoplanet atmospheres. In our case, we apply this technique to visually confirm the presence of neutron-capture elements within the *Gaia*-RVS wavelength span by taking a weighted average of all lines predicted to be present within the spectrograph's wavelength range:

$$c(v) = \sum_{i=0}^{N} x_i \hat{T}_i(v) \tag{10}$$

Here, $x_i$ refers to each spectral pixel, $\hat{T}_i(v)$ are the sets of statistical weights used to compute the average, and $v$ is the Doppler velocity. This equation calculates $c(v)$, the cross-correlation function that represents the weighted average of spectral lines as a function of Doppler velocities. In the case of an exoplanet transit, Doppler shift is necessary as the planet's signal shifts throughout the observation. For stars, the signal will be located at an approximate Doppler shift of 0 km $s^{-1}$, as each RVS spectrum is shifted to the stellar rest frame. The set of statistical weights $\hat{T}_i(v)$ is obtained by computing the expected line positions and depths for a particular star in the data set, *Gaia* ID: 6082274849118517248. For this star, we produce synthetic spectra using the local thermodynamic equilibrium (LTE) stellar line analysis program MOOG[4] (C. Sneden 1973) coupled with one-dimensional LTE model atmospheres from the F. Castelli & R. L. Kurucz (2003) grid. The stellar parameters derived from *The Cannon* are adopted to construct the stellar atmospheric model necessary for the synthesis – i.e. $T_{\rm eff} = 3954$ K, $\log g = 1.16$ dex, [Fe/H] $= -0.7$ dex in this case. In the synthesis, we adjust the line profile to reproduce the resolution of *Gaia*-RVS ($R$∼11 500) and the abundances of Zr, Ce, and Nd for those predicted by *The Cannon*: [Zr/Fe] = 0.27 dex, [Ce/Fe] = 0.31 dex, [Nd/Fe] = 0.81 dex for this star. The line lists used to produce the synthetic spectra in the *Gaia*-RVS wavelength range are generated from LINEMAKE[5] (V. M. Placco et al. 2021) and shown in Table D1. The Ca II triplet lines are also included as a control sample, as explained later in this section.

We now convert the constructed synthetic spectrum to a set of weights using the following formula:

$$\hat{T}_i(v) = \frac{T_i(v)}{\Sigma T_i(v)}. \tag{11}$$

This transformation ensures that the total sum of the templates equals one, with the largest weights allocated to the line positions.

To verify the presence of neutron-capture elements, we cross-correlated each template with a sub-sample of 2000 *Gaia*-RVS spectra randomly selected from a sample of stars with reduced $\chi^2 \approx 1$ between the *The Cannon* model estimates of flux and the observed flux. We also searched for ionized Ca as a proof-of-concept for the application of the technique to our science case.

We cross-correlated over a range of $-50$ to 50 km s$^{-1}$ in 1 km s$^{-1}$ intervals. It is important to note that *Gaia*-RVS spectra are known to have a fluctuating continuum level and occasionally contains aberrant data points. To normalize the cross-correlation function (CCF), we first masked the central region from $-10$ to $+10$ km s$^{-1}$ to avoid the main absorption peak biasing the baseline. The remaining CCF was smoothed using a one-dimensional Gaussian filter to define the continuum, and the entire CCF was then divided by this smoothed profile. This rectified and flattened the continuum, enabling us to resolve the absorption features of the element under examination and allowing reliable detection even for weak lines. Each normalized CCF was then averaged over all stars to produce a one-dimensional cross-correlation function representing the total average line depth across the stars within our sample; for a more detailed description of the method used in this work, see N. W. Borsato et al. (2023).

To assess the robustness of the technique, we propagated the flux uncertainties of the RVS spectra through 1000 Monte Carlo realisations. In each iteration, the flux of each stellar spectrum was randomly sampled according to its associated measurement errors, assuming Gaussian noise, and the corresponding cross-correlation functions were recomputed and averaged. This procedure yielded 1000 realizations of the mean CCF, providing a rigorous estimate of the uncertainty and significance of the recovered absorption features. Our results are shown in Fig. 5, which displays the average CCF with its 1$\sigma$ significance region. As expected, the ionized Ca II peak shows a symmetrical curve and recovers the average line shape of the absorption to a high degree of significance, as indicated by the grey region in this figure; a less significant signal from the recovered line would produce a larger grey area throughout the absorption feature in that figure. Zr, Ce, and Nd are well resolved, but at lower significance, suggesting that the flux probed by these lines is likely weak within the noise of the spectra and was only resolvable by combining lines. Nevertheless, the presence of peaks in Zr, Ce, and Nd (as well as Ca) indicates that typical RVS spectra contain measurable flux data for these elements. Combined with the pseudo-differential analysis, the cross-correlation test supports the interpretation that the *Gaia*-RVS spectra contain measurable neutron-capture information that can contribute to the *The Cannon*-inferred abundances. The resulting catalogue contains stellar label uncertainties for each star and element, allowing users to assess detection significance on a star-by-star basis as appropriate for their applications.

[4]Available at https://www.as.utexas.edu/~chris/moog.html

[5]https://github.com/vmplacco/linemake

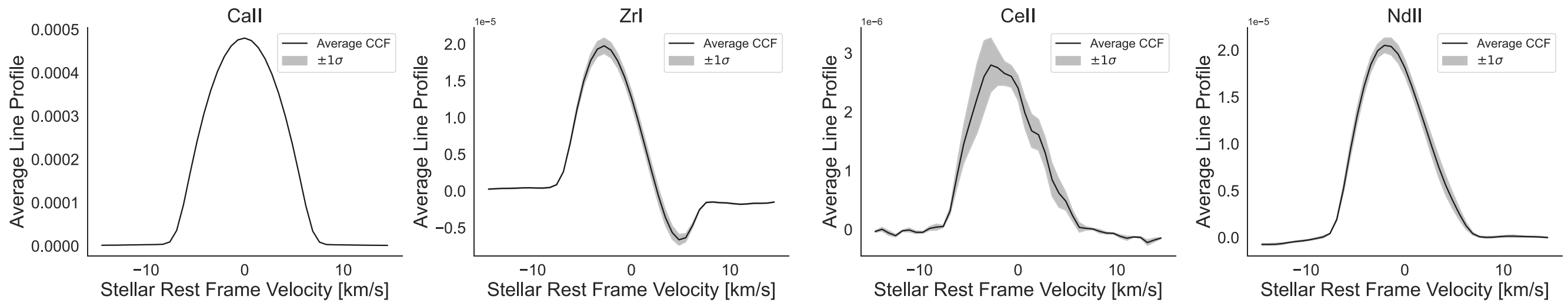

**Figure 5.** Cross-correlation detections of Ca II, Zr I, Ce II, and Nd II. The cross-correlation signal is shown as a black line with its 1$\sigma$ confidence interval marked as a grey region.

## 5 CHEMODYNAMICAL IDENTIFICATION OF GAIA–SAUSAGE–ENCELADUS (GSE) MEMBERS

We now turn to assessing the potential of our methods for tracing the chemodynamical evolution of the Milky Way by applying our results to a specific case – namely the identification of stars associated with *Gaia*-Sausage-Enceladus (GSE), a prominent accreted structure in the Milky Way's stellar halo. In particular, we test whether stars selected solely on the basis of their chemical abundance patterns – as predicted by *The Cannon* – can be validated as probable GSE members based on their kinematic properties.

To compute the kinematic quantities of the stars and enable comparison with earlier GSE kinematic studies (particularly the analysis in D. K. Feuillet et al. 2021), we adopted the `MWPotential2014` Galactic potential model from J. Bovy & H.-W. Rix (2013) and J. Bovy (2015), assuming the solar position to be $R_\odot = 8$ kpc and $Z_\odot = 20.8$ pc. Corrections for the solar motion relative to the Local Standard of Rest were applied using the values $U_\odot = 11.1\,\mathrm{km\,s^{-1}}$, $V_\odot = 12.24\,\mathrm{km\,s^{-1}}$, and $W_\odot = 7.25\,\mathrm{km\,s^{-1}}$ (M. Bennett & J. Bovy 2019). These corrections were used to transform heliocentric velocities to the Galactocentric frame, after which orbits were integrated and kinematic quantities were computed in the `MWPotential2014`, specifically the angular momentum along the *z*-axis ($L_z$), the radial action ($\sqrt{J_R}$), and the total orbital energy ($E$). To construct a validation set, we cross-matched our sample of 314 010 stars (satisfying `flag_cannon = 0`) against the sample of 679 kinematically selected GSE members from D. K. Feuillet et al. (2021), where we found 86 probable GSE members common to both samples.

### 5.1 GSE membership probabilistic model

With the overlapping sample of 86 likely GSE members, we produced an empirical model based on the measured *The Cannon* abundance labels to predict GSE membership. To achieve this, we designed a probabilistic forward model, coupled with a Markov chain Monte Carlo (MCMC) sampler, which assigns a probability for GSE membership, trained on the abundance patterns of the 86 GSE stars obtained from D. K. Feuillet et al. (2021).

In addition to the training sample, we included 2000 stars randomly selected from the remaining non-overlapping *Gaia*-RVS sample. This ensured that the MCMC model would be able to distinguish GSE members from stars not identified in D. K. Feuillet et al. (2021). This comparison sample was chosen to provide a representative background population against which the abundance patterns of the known GSE members could be distinguished. We then used this model to find additional members using the respective stellar labels predicted by *The Cannon*. We adopted a logistic function from D. R. Cox (1958) to predict stellar GSE membership from the *Gaia*-RVS sample. The regression function used for this purpose has the following form:

$$P(Y=1 \mid X) = \frac{1}{1+e^{-(\beta_0+\beta_1 X_1+\beta_2 X_2+\dots+\beta_n X_n)}}. \quad (12)$$

Here, $P(Y=1 \mid X)$ is the probability that the dependent variable $Y$ equals 1 given the input vector $X$, $\beta_0$ is the intercept (bias term), and $\beta_1, \beta_2, \dots, \beta_n$ are the coefficients (weights) associated with the features $X_1, X_2, \dots, X_n$. This logistic equation produces a value between zero and one, enabling the creation of models that flag membership in a specific sub-set. The coefficient values obtained from equation (12) were used to create models indicating GSE membership.

The coefficients selected for regression were the abundance labels [Fe/H] and ([X/Fe] + [Fe/H]) – where X refers to the element in question (X = Ti, Si, Ca, Ni, Zr, Nd and Ce) – to remove any co-linearity between the variables, necessary for producing robust predictor estimates (D. Hosmer, S. Lemeshow & R. Sturdivant 2013). We then calculated the parameters of the logistic function using an MCMC sampler. The model included eight free parameters, one constant term, and eight abundances. All abundance coefficients, including $\beta_{\mathrm{const}}$, $\beta_{\mathrm{Fe/H}}$, $\beta_{\mathrm{Ti/H}}$, $\beta_{\mathrm{Si/H}}$, $\beta_{\mathrm{Ca/H}}$, $\beta_{\mathrm{Ni/H}}$, $\beta_{\mathrm{Zr/H}}$, $\beta_{\mathrm{Nd/H}}$, and $\beta_{\mathrm{Ce/H}}$, were assigned uniform priors $\mathcal{U}(-50, 50)$. This set of broad priors enables an unbiased search for a well-fitting model. We sampled the prior distributions as before, using a No-U-Turn Sampler (M. Betancourt 2017) with the NUMPYRO (E. Bingham et al. 2018; D. Phan, N. Pradhan & M. Jankowiak 2019) and JAX (J. Bradbury et al. 2018) implementations. We sampled the posterior distribution of this model with 100 000 steps, taking 25 000 as burn-in and sampling the distributions for the remaining 75 000 (see Fig. 6). Our results show, in general, good posterior convergence and statistical significance, in addition to reasonable independence for most of the predictor values. The only exception to this is the strong correlation of [Ni/H] with [Fe/H], which is expected since Ni is an Fe-peak element. We present the best-fitting parameter values along with their uncertainties in Table 2.

### 5.2 Interpreting the MCMC framework for GSE candidates

Using the coefficients derived for the model parameters (Table 2), we applied the regression function (equation 12) to each of the 314 010 stars (satisfying `flag_cannon = 0`) in our sample to estimate its GSE membership probability. Our approach is similar to S. Buder et al. (2022), who employed both chemical and dynamical selections using Gaussian Mixture Models to identify candidate GSE stars, demonstrating that purely chemical and purely dynamical selections yield overlapping but not identical samples.

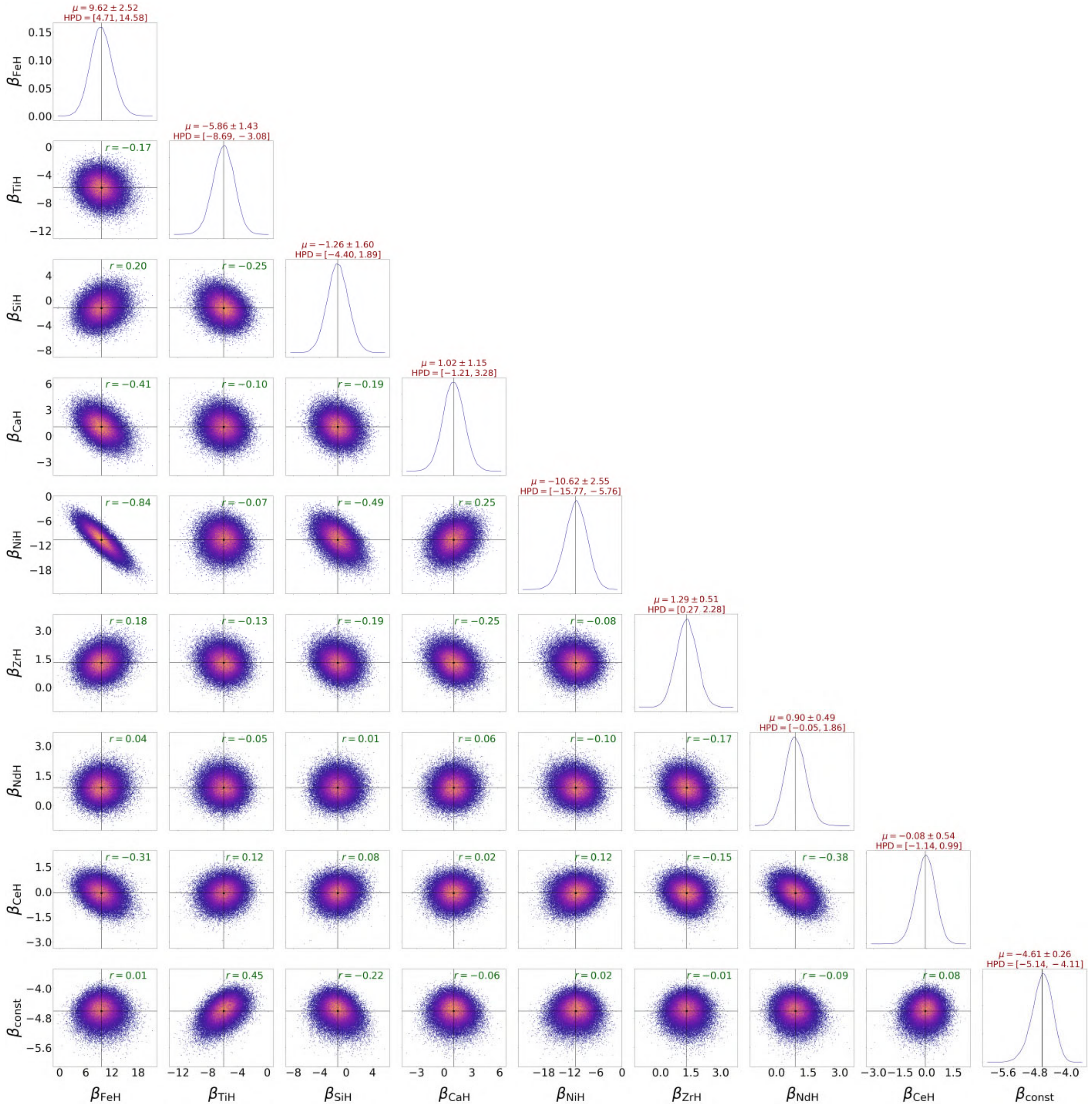

**Figure 6.** Posterior distributions and pairwise marginals for the MCMC model parameters (refer to Section 5.1). Each panel shows samples from the posterior distributions of the abundance ratio coefficients ($\beta$ parameters) inferred from the model. The diagonal panels display the one-dimensional marginal posterior distributions for each parameter, with the mean ($\mu$) and standard deviation ($\sigma$) indicated, together with the 95 per cent Highest Posterior Density (HPD) interval. The off-diagonal hexbin panels illustrate the pairwise relationships between parameters, where colour intensity indicates the density of posterior samples. Annotated values of $\boldsymbol{r}$ denote the Pearson correlation coefficient between each pair of parameters, quantifying the strength and direction of their linear dependence ($r = 1$: perfect positive correlation; $r = -1$: perfect negative correlation; $r = 0$: no linear correlation). Together, these visualisations highlight both the marginal uncertainties and inter-parameter correlations within the inferred posterior distributions.

**Table 2.** Mean and uncertainty values for the coefficients obtained for each stellar abundance parameter used in the MCMC Model (see Section 5.1).

| Parameter | $\beta_{\mathrm{Fe/H}}$ | $\beta_{\mathrm{Ti/H}}$ | $\beta_{\mathrm{Si/H}}$ | $\beta_{\mathrm{Ca/H}}$ | $\beta_{\mathrm{Ni/H}}$ | $\beta_{\mathrm{Zr/H}}$ | $\beta_{\mathrm{Nd/H}}$ | $\beta_{\mathrm{Ce/H}}$ | $\beta_{\mathrm{const}}$ |
|---|---|---|---|---|---|---|---|---|---|
| Mean | 9.62 | −5.86 | −1.26 | 1.02 | −10.62 | 1.29 | 0.90 | −0.08 | −4.61 |
| Uncertainty | 2.52 | 1.43 | 1.60 | 1.15 | 2.55 | 0.51 | 0.49 | 0.54 | 0.26 |

From the sample of 314,010 stars, we identified 3295 potential GSE candidate stars with membership probabilities $P_{\rm mem} >$ 55 per cent, excluding the stars in common with the reference GSE sample from D. K. Feuillet et al. (2021). Table E1 summarizes the membership classification metrics for different probability thresholds, providing the basis for adopting 55 per cent as the selected threshold. To assess the reliability of this threshold, we validated the regression model using the cross-matched MCMC training sample of 86 stars from D. K. Feuillet et al. (2021). At $P_{\rm mem} > 55$ per cent, the model achieves a precision of 69 per cent through maximizing the number of true positives relative to false positives, indicating that nearly two-thirds of stars predicted as GSE members above this threshold are expected to have correct probabilistic classifications. Based on this validation, we selected the 2289 high-confidence candidates ($\sim$ 69 per cent of 3295) as our final sample of probable GSE members for further analysis.

To distinguish these 2289 probable GSE stars from the *in situ* halo and thick-disc populations, we constructed two comparison samples of probable non-GSE stars. The first sample includes *in situ* halo stars, with [Fe/H] $< -0.8$ dex and [Ti/Fe] $> 0.25$ dex, comprising chemically older and $\alpha$-enhanced populations formed within the Milky Way. These stars were further selected to have angular momenta in the range $-1500 < L_z$ (kpc km s$^{-1}$) $< 1500$ and orbital eccentricities $0.5 < e < 0.8$, following kinematic boundaries commonly adopted to separate accreted and *in situ* halo components (see e.g. V. Belokurov et al. 2018; H. Koppelman, A. Helmi & J. Veljanoski 2018; H. H. Koppelman et al. 2019; R. P. Naidu et al. 2020; D. K. Feuillet et al. 2021). The roughly symmetric range around $L_z \approx 0$ kpc km s$^{-1}$ isolates stars on low-angular-momentum, moderately eccentric orbits typical of the *in situ* halo, while excluding the strongly radial GSE debris. For the non-GSE thick-disc sample, we adopted the same chemical criteria as for the *in situ* halo, but required $L_z > 0$ kpc km s$^{-1}$ and $e < 0.5$ to select stars on more circular, prograde, rotationally supported orbits characteristic of the Galactic disc. This kinematic separation increases the likelihood that the comparison populations represent stars formed within the Milky Way, enabling a better assessment of chemical and dynamical distinctions between *in situ* and accreted halo components. We also note that, primarily due to their lower metallicities, GSE stars can be more readily distinguished from the thin-disc population.

Using *The Cannon*-predicted chemical abundances, we examined their pairwise abundance distributions across four distinct stellar samples: (a) the sample of 2289 high-confidence probable GSE members with membership probabilities $P_{\rm mem} > 55$ per cent; (b) the sample of non-GSE halo stars selected based on the criteria outlined previously; (c) the sample of non-GSE thick-disc stars, also defined as described earlier; and (d) the reference sample of the 86 GSE stars common to our catalogue of RVS giants and the kinematically selected GSE sample from D. K. Feuillet et al. (2021) (see Fig. F1 for all the pairwise abundance distributions). Visual inspection of these distributions – subsequently validated using a Random Forest classifier – shows that the abundance ratios [Ca/Ti], [Ti/Ce], and [Nd/Zr] exhibit similar trends in both the predicted GSE sample and the reference GSE sample, while being distinct from those of the thick-disc and halo stellar populations. These differences are further highlighted in Fig. 7, which focuses on the pairwise correlations of the selected abundance ratios. The distributions of peak population densities differ notably across the samples, emphasizing the use of these abundance ratios for identifying probable GSE candidates.

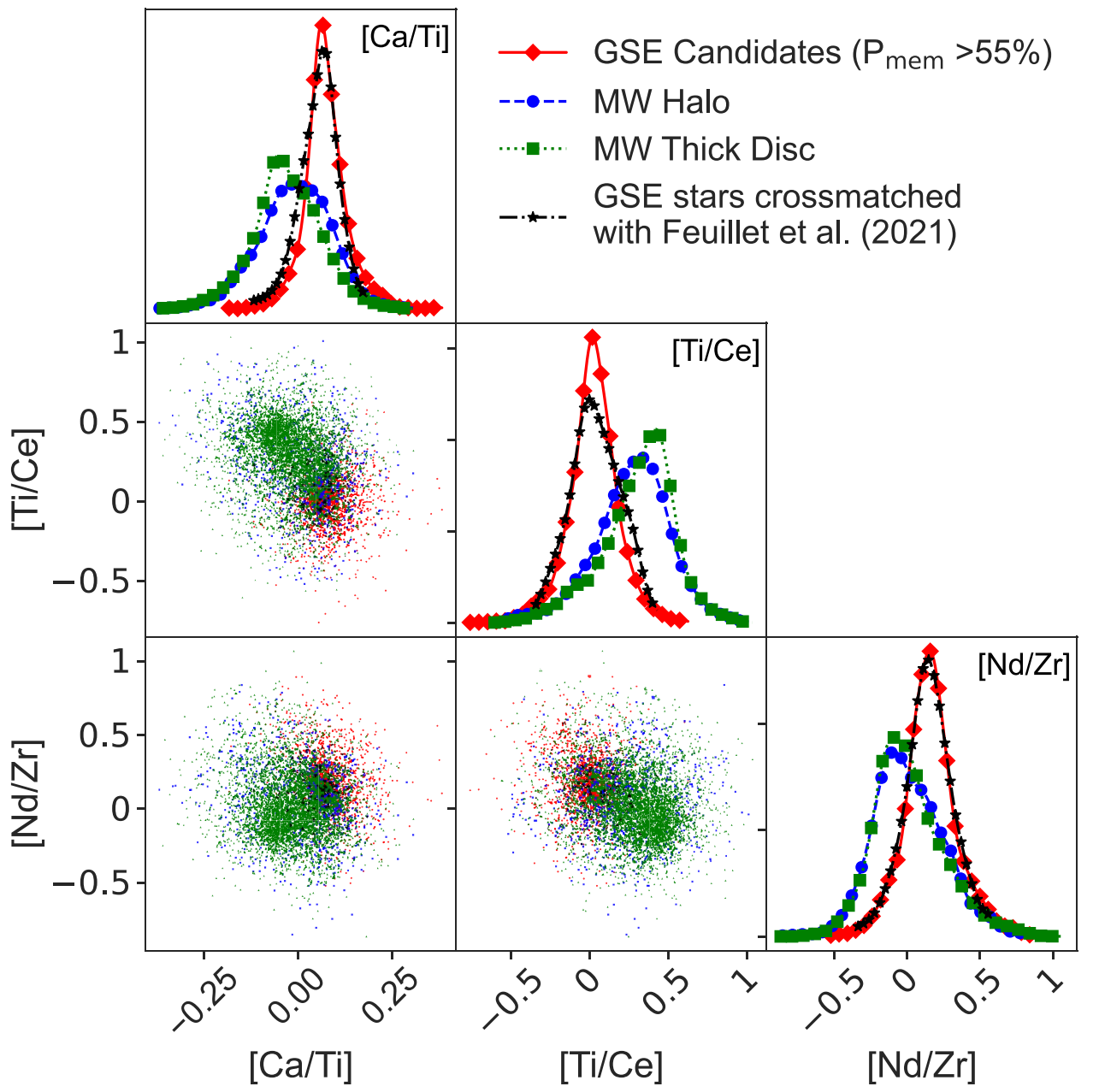

**Figure 7.** Distributions and pairwise correlations of the abundance ratios – [Ca/Ti], [Ti/Ce], and [Nd/Zr] – for four distinct stellar populations: 2289 probable GSE candidate stars with $P_{\rm mem} > 55$ per cent (red); halo stars (blue), thick disc stars (green), and a cross-match of 86 GSE stars from D. K. Feuillet et al. (2021) with our catalogue of 314 010 *Gaia*-RVS stars (black) satisfying `flag_cannon = 0`. Diagonal panels display kernel density estimates of the one-dimensional distributions, while the lower off-diagonal panels show corresponding two-dimensional scatter plots. The relative chemical distributions for these the ratios ([Ca/Ti], [Ti/Ce], and [Nd/Zr]) highlight distinct signatures between accreted (GSE) and *in situ* (halo and thick disc) populations, supporting the use of these abundance ratios as discriminants of stellar origin in the Milky Way halo.

We also show kernel density estimates (KDEs) of the abundance ratios [Ca/Ti], [Ti/Ce], and [Nd/Zr] for our sample of 2289 GSE candidate stars and the chemodynamically selected comparison sample of non-GSE halo stars in Fig. 8. For each abundance ratio, we indicate the location of the GSE KDE peak and its associated 1$\sigma$ interval, defined by the 16th and 84th percentiles of the GSE distribution. Within these GSE 1$\sigma$ intervals, we find that 376 ($\sim$29 per cent), 283 ($\sim$22 per cent), and 462 ($\sim$36 per cent) of the halo stars fall within the 1$\sigma$ intervals for [Ca/Ti], [Ti/Ce], and [Nd/Zr], respectively. These abundance ratios show intriguing differences among individual $\alpha$- and neutron-capture elements. The higher [Ca/Ti] ratios observed for our GSE candidates relative to the *in situ* halo and thick-disc populations may indicate subtle variations within the $\alpha$-element group (D. Horta & M. K. Ness 2025). Similarly, the lower [Ti/Ce] and slightly higher [Nd/Zr] in our sample may reflect intra-process variations within the neutron-capture domain, arising from differing contributions of slow- and rapid-process enrichment channels. While D. Horta & M. K. Ness (2025) focused on hydrostatic versus explosive $\alpha$-element production, our results suggest that analogous intra-process variations may also exist among neutron-capture elements in accreted systems such as GSE, highlighting subtle internal differences within element families linked to their star-formation and chemical-enrichment histories, a topic which warrants future exploration.

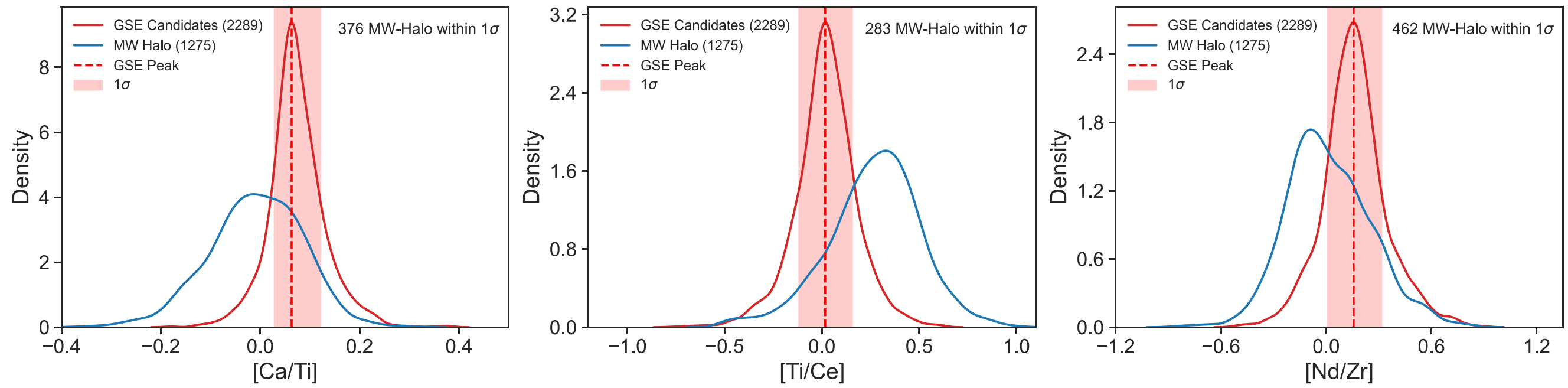

**Figure 8.** Kernel density estimates of the selected chemical abundance ratios – [Ca/Ti], [Ti/Ce], and [Nd/Zr] – for 2289 GSE candidate stars (red), and the Milky Way halo stars (blue) restricted to [Fe/H] $< -0.8$, [Ti/Fe] $> 0.25$ dex, $0.5 < e < 0.8$, and $|L_z| < 1500$ kpc km s$^{-1}$. For each abundance ratio, the peak of the distribution (dashed line) and $1\sigma$ interval (16th–84th percentile, shaded region) are indicated. Annotations in each panel report the number of halo stars falling within the GSE $1\sigma$ range. These comparisons highlight systematic chemical distinctions between the GSE and the *in situ* halo, reflecting different nucleosynthetic enrichment histories.

These patterns are broadly consistent with earlier findings for Galactic stellar populations. For example, P. E. Nissen & W. J. Schuster (2010) identified two distinct halo populations based on [$\alpha$/Fe] ratios, with low-$\alpha$ stars exhibiting signatures of slower star formation and likely accreted origin and high-$\alpha$ stars reflecting rapid *in situ* enrichment. Similarly, C. K. Fishlock et al. (2017) showed that low-$\alpha$ halo stars display characteristic neutron-capture abundance patterns, such as higher [Ba/Y] and lower [Y/Eu], indicative of contributions from low-metallicity AGB stars. Our observed differences in [Ca/Ti], [Ti/Ce], and [Nd/Zr] among GSE candidates versus *in situ* populations are conceptually similar to these studies, emphasizing the diagnostic power of intra-process abundance ratios to distinguish accreted from *in situ* stars.

Our chemical-abundance-based selection of GSE candidates complements alternative approaches employed in the literature. For instance, K. Hawkins et al. (2015) and P. Das et al. (2020) utilized abundance planes such as [Al/Fe] versus [Mg/Mn] to separate accreted from *in situ* populations, while S. Buder et al. (2022) employed [Na/Fe] versus [Mg/Mn] in combination with Gaussian Mixture Models for the same purpose. Although these methods rely on different chemical tracers, they share the underlying principle of using element ratios sensitive to star formation history and nucleosynthetic pathways to distinguish accreted substructures. Hence our approach, based on [Ca/Ti], [Ti/Ce], and [Nd/Zr], provides a complementary perspective, particularly highlighting intra-process differences among $\alpha$- and neutron-capture elements.

To further investigate the overlap of our sample of 2289 GSE candidates with the dynamically selected GSE sample of D. K. Feuillet et al. (2021), we applied kinematic criteria of $-500 \leq L_z$ (kpc km s$^{-1}$) $\leq 500$ and $30 \leq \sqrt{J_r}$ (kpc$^{1/2}$ km$^{1/2}$ s$^{-1/2}$) $\leq 55$. Applying these cuts to the initial sample resulted in a sample selection of 286 stars that satisfy both our probabilistic chemical selection and the dynamical criteria.

Fig. 9 shows the normalized distributions of eight elemental abundance ratios ([Fe/H], [Ti/Fe], [Ca/Fe], [Ni/Fe], [Si/Fe], [Nd/Fe], [Zr/Fe], and [Ce/Fe]) for four samples: (i) the chemically selected high-confidence GSE candidates ($P_{\rm mem} > 55$ per cent, 2289 stars), (ii) the sub-set of 286 stars also satisfying dynamical criteria [$-500 \leq L_z$ (kpc km s$^{-1}$) $\leq 500$ and $30 \leq \sqrt{J_r}$ (kpc$^{1/2}$ km$^{1/2}$ s$^{-1/2}$) $\leq 55$], and (iii–iv) GALAH DR3 and DR4 abundances for the GSE sample from S. Buder et al. (2022) (using their recommended membership $>$ 45 per cent, based on chemical selection). The coloured lines represent the relative distributions of each sample; the inset legend in each panel reports the median and 16th–84th percentile spreads (dex). We also calculate the median abundances for all the elements analysed in Table 3. We show that the 286 chemodynamically selected stars exhibit median abundances that closely match those of the full chemical sample of 2289 GSE candidates, demonstrating that the kinematic cut does not introduce notable bias in the overall chemical patterns. The median abundances from GALAH DR4 are generally consistent with both the chemically selected and the chemodynamically selected samples, reflecting the fact that *The Cannon* was trained using GALAH DR4 labels as the reference in the model. In contrast, the GALAH DR3 abundances show systematic offsets and exhibit broader distributions or slightly different medians for the $\alpha$- ([Ca/Fe], [Ti/Fe]) and neutron-capture elements ([Nd/Fe], [Zr/Fe], [Ce/Fe]). These differences likely reflect the variations in label calibration and the analysis methodologies adopted in GALAH DR3 versus DR4.

For the sample of 286 GSE candidates, consisting exclusively of RVS giant stars, *The Cannon*-predicted metallicity is obtained as [Fe/H] $= -1.56^{+0.23}_{-0.24}$ dex. We note that this result reflects a more metal-poor population compared to previously reported literature values for GSE, such as $\sim(-1.17 \pm 0.34)$ dex in D. K. Feuillet et al. (2020), $\sim(-1.15)$ dex in D. K. Feuillet et al. (2021) and $(-1.15^{-0.24}_{-0.33})$ dex in R. P. Naidu et al. (2020). A possible explanation for this metallicity difference is the exclusive use of giant stars in our sample, as GSE membership probabilities may be higher due to better chemical distinctions from the *in situ* halo field at low metallicity; incorporating GSE dwarfs in the sample would likely yield a higher mean metallicity. This can be seen in fig. 3 of H. Ernandes et al. (2024), where GSE giants display a more metal-poor distribution relative to the GSE dwarfs within the same primary sample.

To visualize how our chemically selected GSE candidates occupy dynamical space, Fig. 10 presents the distribution of the 2289 probable GSE stars in the $L_z$–$E$ and $L_z$–$\sqrt{J_R}$ planes. The majority of these stars cluster around low angular momentum ($L_z \approx 0$) and moderate-to-high $\sqrt{J_R}$ and eccentricities, consistent with the highly radial, weakly rotating orbits characteristic of GSE debris. This behaviour is in agreement with previous studies showing that the GSE progenitor was accreted on a predominantly radial orbit, leaving behind stars with little net rotation and elevated orbital energies (V. Belokurov et al. 2018; D. K. Feuillet et al. 2021; R. P. Naidu et al. 2021). Note that, as our selection is based purely on chemical abundances, there are stars distributed across a fairly broad range of dynamical space. While a substantial fraction lies within the canonical GSE region defined by low $L_z$ and high radial action, a subset of stars extends beyond these bounds, potentially reflecting the natural diversity in the orbital properties of the accreted debris, as well as perhaps some degree of contamination by stars of different origins. If intrinsic to GSE stars, the spread could result from variations in the progenitor's structure, the range of energies and angular momenta at infall, and subsequent dynamical evolution within the Milky Way potential. Recent reanalyses of the sample from P. E. Nissen & W. J. Schuster (2010) provide further evidence for GSE debris spanning a broad kinematic range, with chemical tracers such as [Al/Fe] identifying GSE-like stars across varied orbital energies, and indications of multiple progenitor passages producing a metallicity–$E$ gradient (H. Ernandes et al. 2025; Á.

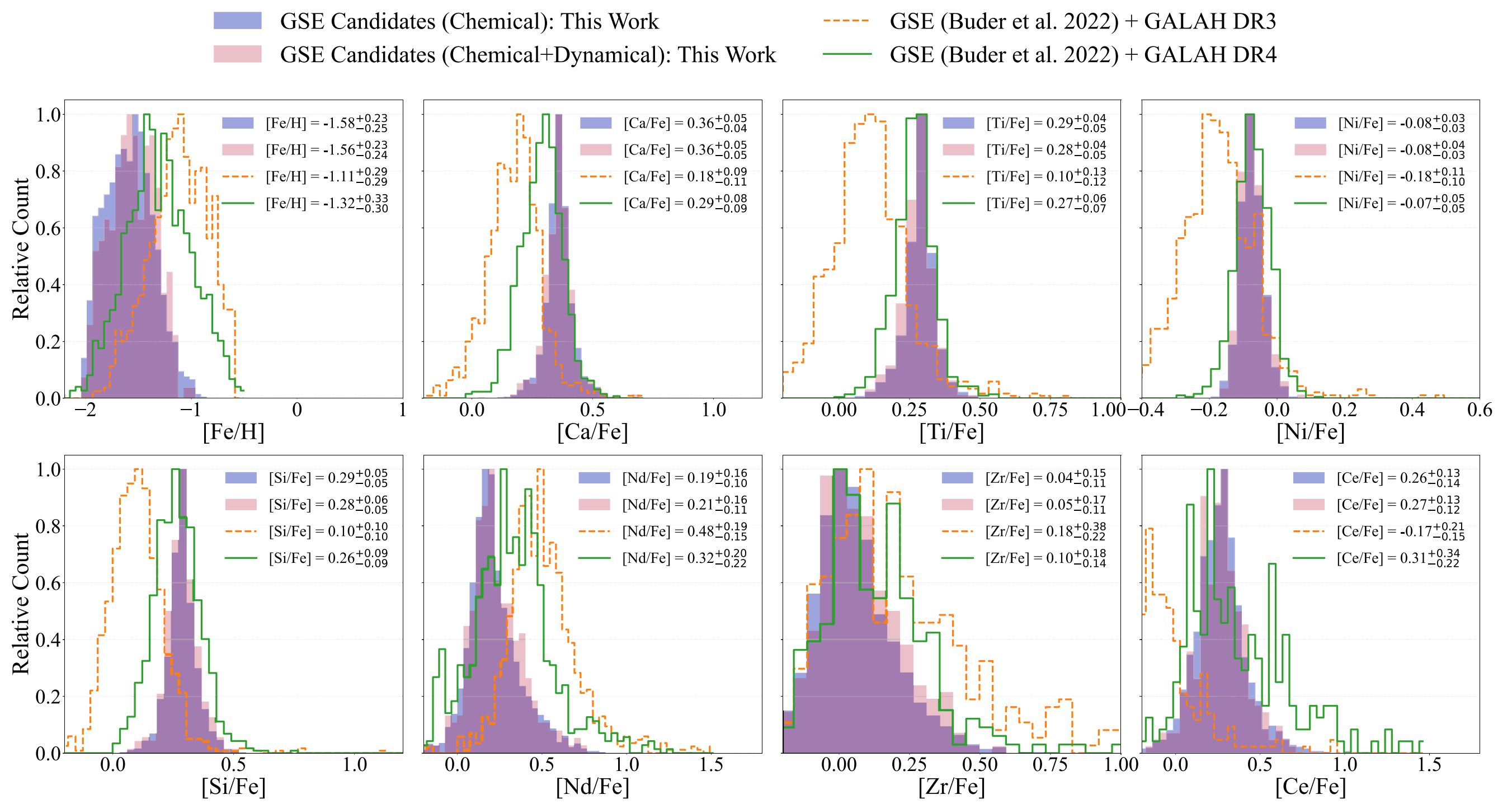

**Figure 9.** Each panel shows the normalized histogram of the abundance ratio (dex) for four samples of GSE stars: 2289 GSE candidates selected chemically in our work ($P_{\rm mem} > 55$ per cent) marked with blue; a sub-set of 286 GSE candidates selected chemically in our sample ($P_{\rm mem} > 55$ per cent) and satisfying the dynamical conditions from D. K. Feuillet et al. (2021) [$-500 \leq L_z$ (kpc km s$^{-1}$) $\leq 500$ and $30 \leq \sqrt{J_r}$ (kpc$^{1/2}$ km$^{1/2}$ s$^{-1/2}$) $\leq 55$] marked with pink; and the GALAH DR3 and DR4 elemental abundances cross-matched with the GSE sample from S. Buder et al. (2022), satisfying their membership criterion > 45 per cent for [Na/Fe] versus [Mg/Mn], marked with orange and green respectively. Coloured lines indicate the relative distribution for each data set. The inset legend in each panel gives the median value with 16th–84th percentile spreads (dex) for each sample within each elemental space. The vertical axis represents the relative frequency, normalized to the maximum of each distribution. The global legend above the panels refers to the data sets.

**Table 3.** Median elemental abundances (dex) with 16th–84th percentile spreads in the four GSE samples: (a) 2289 chemically selected GSE candidates ($P_{\rm mem} > 55$ per cent); (b) 286 GSE candidates selected using both chemical and dynamical criteria [$P_{\rm mem} > 55$ per cent, $-500 \leq L_z$ (kpc km s$^{-1}$) $\leq 500$, and $30 \leq \sqrt{J_r}$ (kpc$^{1/2}$ km$^{1/2}$ s$^{-1/2}$) $\leq 55$]; (c) and (d) denote GSE stars identified by S. Buder et al. (2022) based on their [Na/Fe]–[Mg/Mn] distribution (membership probability > 45 per cent), with elemental abundances taken from GALAH DR3 and DR4, respectively.

| Element | GSE candidates (Chemical)[a] | GSE candidates (Chemical + Dynamical)[b] | GALAH DR3 GSE[c] | GALAH DR4 GSE[d] |
|---|---|---|---|---|
| [Fe/H] | $-1.58^{+0.23}_{-0.25}$ | $-1.56^{+0.23}_{-0.24}$ | $-1.11^{+0.29}_{-0.29}$ | $-1.32^{+0.33}_{-0.30}$ |
| [Ca/Fe] | $0.36^{+0.05}_{-0.04}$ | $0.36^{+0.05}_{-0.05}$ | $0.18^{+0.09}_{-0.11}$ | $0.29^{+0.08}_{-0.09}$ |
| [Ti/Fe] | $0.29^{+0.04}_{-0.05}$ | $0.28^{+0.04}_{-0.05}$ | $0.10^{+0.13}_{-0.12}$ | $0.27^{+0.06}_{-0.07}$ |
| [Ni/Fe] | $-0.08^{+0.03}_{-0.03}$ | $-0.08^{+0.04}_{-0.03}$ | $-0.18^{+0.11}_{-0.10}$ | $-0.07^{+0.05}_{-0.05}$ |
| [Si/Fe] | $0.29^{+0.05}_{-0.05}$ | $0.28^{+0.06}_{-0.05}$ | $0.10^{+0.10}_{-0.10}$ | $0.26^{+0.09}_{-0.09}$ |
| [Nd/Fe] | $0.19^{+0.16}_{-0.10}$ | $0.21^{+0.16}_{-0.11}$ | $0.48^{+0.19}_{-0.15}$ | $0.32^{+0.20}_{-0.22}$ |
| [Zr/Fe] | $0.04^{+0.15}_{-0.11}$ | $0.05^{+0.17}_{-0.11}$ | $0.18^{+0.38}_{-0.22}$ | $0.10^{+0.18}_{-0.14}$ |
| [Ce/Fe] | $0.26^{+0.13}_{-0.14}$ | $0.27^{+0.13}_{-0.12}$ | $-0.17^{+0.21}_{-0.15}$ | $0.31^{+0.34}_{-0.22}$ |

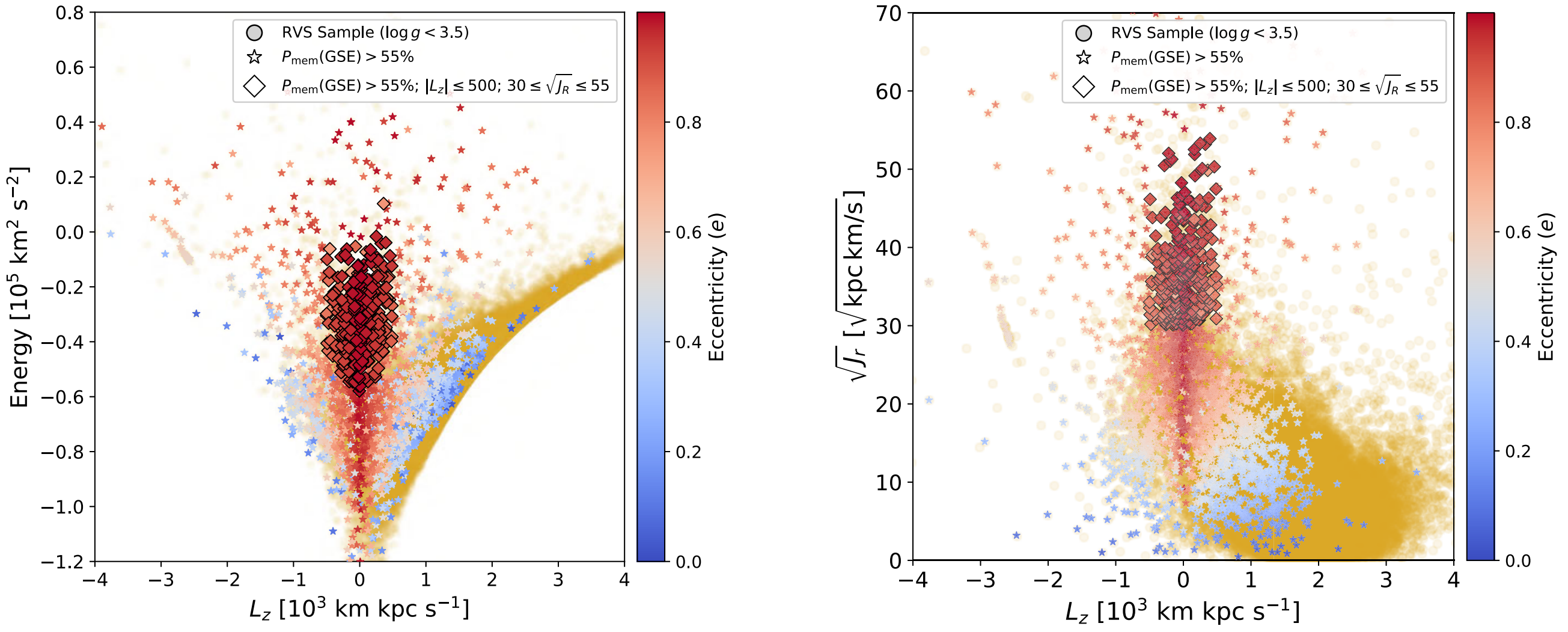

**Figure 10.** Kinematic distribution of red giant stars analysed with *The Cannon*. **Left:** Total orbital energy ($E$) versus angular momentum ($L_z$). The golden coloured-points show the RVS sample of 314 010 RVS giants (flag_cannon = 0). Star-shaped markers denote probable GSE members with $P_{\rm mem} >$ 55 per cent, selected using the abundance-based MCMC model. Diamond-shaped markers show the sub-set of stars that also satisfy the dynamical cuts from D. K. Feuillet et al. (2021), namely $|L_z| \le 500$ kpc km s$^{-1}$ and $30 \le \sqrt{J_R}$(kpc$^{1/2}$ km$^{1/2}$ s$^{-1/2}$) $\le 55$. **Right:** Radial action ($\sqrt{J_R}$) versus $L_z$ for the same sample of stars. The various markers denote the same objects as it was in the left panel. Colour bars indicate eccentricity ($e$) for the probable GSE candidates only.

Skúladóttir et al. 2025; L. Berni et al. 2026). This supports the interpretation that the kinematic diversity in our chemically selected sample is largely intrinsic to GSE. Overall, these results suggest that the chemical selection broadly traces the GSE debris population, while dynamical refinement identifies a subset that is more tightly concentrated in the traditionally defined GSE region.

## 6 CONCLUSIONS

Our study demonstrates the ability of *The Cannon* to transfer high-resolution GALAH DR4 stellar labels ($R \approx$ 28 000) to relatively lower-resolution *Gaia* DR3 RVS spectra ($R \approx$ 11 500), enabling inference of stellar parameters ($T_{\rm eff}$, log g, and $v \sin i$) and a broad range of elemental abundances, including [Fe/H], [Ca/Fe], [Si/Fe], [Ni/Fe], and [Ti/Fe], for a sample of 357 415 RVS giant stars (using $\log g < 3.5$ dex from the catalogue in P. B. Das et al. 2025). Notably, the model also predicts neutron-capture element abundances such as [Zr/Fe], [Ce/Fe], and [Nd/Fe] with low scatter. 12-fold cross-validation on the training set confirms internal consistency, with RMSE values of ~70 K ($T_{\rm eff}$), ~0.14 dex (log g), $\lesssim$0.10 dex for most abundances and ~0.15 dex for neutron-capture elements. Of these, 314 010 stars satisfy flag_cannon = 0, representing a high-fidelity subsample within the training label-space bounds.

To independently assess the detectability of neutron-capture features in the RVS spectra, we carried out a pseudo-differential analysis of two stars with similar stellar parameters but different [Nd/Fe], illustrating the small (but noticeable) flux variations in the Nd lines. We also applied a multiline cross-correlation technique to a representative sub-sample of 2000 RVS giants satisfying a reduced chi-squared value ($\chi^2$) $\approx$ 1 between the *The Cannon* model estimates of flux and the observed RVS spectral flux. This latter approach showed well-defined peaks in the cross-correlation functions for Zr, Ce, and Nd absorption features, supporting the presence of measurable abundance information in the spectral range. Combined together, these results support the reliability of *The Cannon*'s abundance predictions for the neutron-capture elements.

Applying the results of our analysis to the stellar populations of the Milky Way, we used an empirical, abundance-based framework optimized with an MCMC sampler to identify probable GSE members. The training sample consisted of 86 stars overlapping between our catalogue and the kinematically selected GSE sample from D. K. Feuillet et al. (2021), supplemented by a sample of 2000 randomly selected non-GSE stars to ensure distinction from *in situ* populations. This model produced membership probabilities for the 314 010 RVS giants (`flag_cannon = 0`), identifying 3295 stars exceeding a probability threshold of 55 per cent, after excluding the stars cross-matched with GSE reference sample from D. K. Feuillet et al. (2021). Accounting for the estimated precision of approximately 69 per cent at this threshold (i.e. considering both true and false positives), we selected 2289 stars with the highest probabilities ($\approx$ 69 per cent of 3295) as our high-confidence probable GSE candidates.

**Table 4.** Median dynamical properties with 16th and 84th percentile uncertainties for the chemically and chemodynamically selected GSE candidates.

| Parameter | GSE candidates (Chemical)[a] | GSE candidates (Chemical + Dynamical)[b] |
|---|---|---|
| $\sqrt{J_R}$ (kpc$^{1/2}$ km$^{1/2}$ s$^{-1/2}$) | $19.37^{-8.98}_{+14.02}$ | $35.82^{-4.27}_{+7.39}$ |
| $L_x$ (kpc km s$^{-1}$) | $36.77^{-316.20}_{+330.20}$ | $11.58^{-225.12}_{+232.74}$ |
| $L_y$ (kpc km s$^{-1}$) | $-35.24^{-533.26}_{+545.86}$ | $-19.57^{-572.98}_{+507.89}$ |
| $L_z$ (kpc km s$^{-1}$) | $170.56^{-660.80}_{+800.08}$ | $-12.72^{-238.50}_{+279.61}$ |
| $E$ ($10^5$ km$^2$ s$^{-2}$) | $-0.57^{-0.19}_{+0.29}$ | $-0.33^{-0.11}_{+0.15}$ |
| $V_R$ (km s$^{-1}$) | $1.55^{-154.64}_{+158.05}$ | $85.99^{-378.11}_{+204.34}$ |
| $V_\phi$ (km s$^{-1}$) | $28.21^{-107.40}_{+122.94}$ | $-1.66^{-29.67}_{+37.83}$ |
| $V_z$ (km s$^{-1}$) | $-4.91^{-96.87}_{+92.06}$ | $2.43^{-104.03}_{+110.94}$ |
| $e$ (eccentricity) | $0.70^{-0.33}_{+0.23}$ | $0.95^{-0.05}_{+0.04}$ |

[a] GSE Candidates with $P_{\rm mem} >$ 55 per cent (2289 stars).
[b] GSE Candidates with Pmem > 55 per cent and satisfying $|L_z| \le$ 500 kpc km s$^{-1}$ and $30 \le \sqrt{J_R}$ (kpc$^{1/2}$ km$^{1/2}$ s$^{-1/2}$) $\le 55$ (286 stars).

On comparing the chemical abundance ratios of these 2289 probable GSE candidates to that of *in situ* Milky Way halo and thick-disc comparison samples, the abundance ratios [Ca/Ti], [Ti/Ce], and [Nd/Zr] showed distinct distributions between the GSE candidates and *in situ* populations, highlighting differences in nucleosynthetic histories. This was further validated by a Random Forest classifier, confirming that these abundance ratios can potentially separate the chemically selected GSE stars from *in situ* halo and disc stars. These patterns are conceptually consistent with previous studies of accreted populations and halo chemistry, such as P. E. Nissen & W. J. Schuster (2010) and C. K. Fishlock et al. (2017), and are broadly aligned with previous chemodynamical selections of GSE stars (S. Buder et al. 2022).

We further examined the dynamical properties of these 2289 GSE candidates using the MWPotential2014 Galactic potential model from J. Bovy & H.-W. Rix (2013) and J. Bovy (2015). The majority of the 2289 GSE candidates occupy around low angular momentum ($L_z \approx 0$) and moderate-to-high radial action ($\sqrt{J_R}$), corresponding to the highly radial, weakly rotating orbits characteristic of GSE debris. This is in agreement with previous studies suggesting that the GSE progenitor was accreted on a predominantly radial orbit, leaving behind stars with little net rotation and elevated orbital energies (V. Belokurov et al. 2018; R. P. Naidu et al. 2021). The stars are distributed across a broad dynamical range, potentially reflecting natural diversity in the progenitor's orbital energy, angular momentum, and subsequent evolution in the Milky Way potential, as well as possibly indicating some degree of contamination from stars having different origins. Applying dynamical cuts from D. K. Feuillet et al. (2021) [$-500 \leq L_z$ (kpc km s$^{-1}$) $\leq 500$ and $30 \leq \sqrt{J_R}$ (kpc$^{1/2}$ km$^{1/2}$ s$^{-1/2}$) $\leq 55$] identifies 286 stars within the canonical GSE region, showing that the purely chemical selection broadly traces the debris, while the dynamical selection defines a more concentrated sub-set. The mean predicted metallicity of this dynamically selected subset is obtained as [Fe/H] = $-1.56^{+0.23}_{-0.24}$ dex, a relatively more metal poor population as compared to the exsiting literature sources (D. K. Feuillet et al. 2020, 2021; R. P. Naidu et al. 2020). This is likely due to the exclusive inclusion of RVS giants in our sample which yields more metal poor-population. This can also be seen in fig. 3 of H. Ernandes et al. (2024), where the GSE giants displayed a more metal-poor distribution as compared to the GSE dwarf population. Table 4 presents the kinematic properties calculated for the samples of chemically (2289) and chemodynamically (286) selected GSE candidates.

Overall, our results demonstrate that combining data-driven chemical abundance inference with probabilistic classification and dynamical validation provides a robust and scalable framework for identifying accreted stellar populations. *The Cannon*, trained on high-resolution reference data, can reliably infer both $\alpha$- and neutron-capture element abundances for large samples of moderate-resolution spectra when complemented by appropriate statistical tests. Chemical selection criteria effectively identify probable GSE stars, while dynamical comparisons provide the kinematic context necessary to interpret the orbital properties of these accreted populations. Although these abundance estimates are robust at the population level, results for individual stars (particularly those near the edges of the parameter space or with lower SNR) may carry higher uncertainties.

The resulting chemical patterns of the GSE candidates – involving abundance ratios of a combination of alpha and neutron-capture elements, in our case [Ca/Ti], [Ti/Ce] and [Nd/Zr] – provide a statistically robust view of the stellar populations accreted during this merger event. While other elemental ratios have been used to distinguish between accreted and *in situ* populations (e.g. K. Hawkins et al. 2015; P. Das et al. 2020; S. Buder et al. 2022; H. Ernandes et al. 2025), the addition of neutron-capture elements provides additional constraints on the detailed chemical evolution of GSE. This approach can be readily extended to characterize other halo substructures both in *Gaia*-RVS data (e.g. Das et al. in preparation) and in upcoming large-scale spectroscopic surveys. A key question to be addressed in future work is whether the specific elemental abundance ratios we find to be characteristic of GSE are distinct from those associated with other accretion events. With the approaching advent of *Gaia* DR4, which will include over 200 million RVS spectra – more than a hundred-fold increase over DR3 – this methodology will enable the chemodynamical mapping of a significant sample of the Galaxy's stellar populations, allowing us to trace the Milky Way's accretion history in unprecedented detail, and thereby help in disentangling the complex process of Galactic assembly.

## ACKNOWLEDGEMENTS

PBD is supported by the Australian Commonwealth Government Research Training Program (RTP) Scholarship (https://doi.org/10.82133/C42F-K220). PBD acknowledges support from the Astronomical Society of Australia (ASA) Student Travel Grant 2025 (https://asa.astronomy.org.au/prizes_and-grants/the-student-travel-grant/), the Macquarie University Postgraduate Research Fund (PGRF) 2025, and the European Southern Observatory (ESO) Garching Early Career Scientific Visitor Programme 2025. Parts of this research were supported by the Australian Research Council (ARC) Centre of Excellence for All Sky Astrophysics in 3 Dimensions (ASTRO-3D), through project number CE170100013.

AMG, DBZ, GFL, and SLM acknowledge support from the ARC through Discovery Project DP220102254. DF acknowledges funding from the Swedish Research Council grant 2022–03274. TN acknowledges support from the Knut and Alice Wallenberg Foundation. SLM acknowledges support from the UNSW Scientia Fellows Program. DS is supported by the Australian Research Council Discovery Project DP250104267. PBD thanks Anish Amarsi and Madeleine McKenzie for helpful discussions related to this work. The authors also thank the anonymous reviewer for the valuable comments and suggestions.

This work has used the Fourth Data Release of the GALAH Survey. The GALAH survey is based on data acquired through the Australian Astronomical Observatory, under the programmes: A/2013B/13 (The GALAH pilot survey); A/2014A/25, A/2015A/19, A2017A/18 (The GALAH survey, Phase 1); A2018A/18 (Open clusters with HERMES); A2019A/1 (Hierarchical star formation in Ori OB1); A2019A/15 (The GALAH survey, Phase 2); A/2015B/19, A/2016A/22, A/2016B/10, A/2017B/16, A/2018B/15 (The HERMES-*TESS* programme); and A/2015A/3, A/2015B/1, A/2015B/19, A/2016A/22, A/2016B/12, A/2017A/14 (The HERMES *K2*-follow-up programme). We acknowledge the traditional owners of the land on which the Anglo-Australian Telescope stands, the Gamilaraay people, and pay our respects to elders past and present. This paper includes data that have been provided by AAO Data Central (https://datacentral.org.au/).

This work has made use of the data from the European Space Agency (ESA) mission *Gaia* (https://www.cosmos.esa.int/gaia), processed by the *Gaia* Data Processing and Analysis Consortium (DPAC, https://www.cosmos.esa.int/web/gaia/dpac/consortium). Funding for the DPAC has been provided by national institutions, in particular the institutions participating in the *Gaia* Multilateral Agreement.

This paper also utilized NASA's Astrophysics Data System (ADS) bibliographic services, as well as the open-source PYTHON packages ASTROPY (http://www.astropy.org) (Astropy Collaboration 2013, 2018, 2022), NUMPY (C. R. Harris et al. 2020), PANDAS (W. McKinney 2010), MATPLOTLIB (J. D. Hunter 2007), and TOPCAT (M. B. Taylor 2005; M. Taylor 2011).

## DATA AVAILABILITY

The data underlying this article is available at the CDS via anonymous FTP at https://cdsarc.u-strasbg.fr (130.79.128.5) or via the VizieR catalogue service at https://cdsarc.cds.unistra.fr/viz-bin/cat/J/MNRAS/550/G1142.

## APPENDIX A: TABLE DESCRIPTION OF *The Cannon* CATALOGUES FOR THE *GAIA*-RVS STARS

Table A1 provides the full column schema of the catalogue produced in this work. The catalogue includes *Gaia* DR3 source identifier, coordinates (RA and Dec), parallax, RVS spectral SNR, *The Cannon*-predicted stellar labels and their corresponding uncertainties, and the quality flags (`flag_cannon`, `flag_FeH` and `flag_XFe`, $X \in$ {Ca, Ti, Si, Ni, Nd, Ce, Zr}), which indicate the reliability of the inferred stellar labels. All abundances are reported in standard logarithmic notation relative to solar values, in units of dex. The reduced-$\chi^2$ statistic (`rchi_sq`) quantifies the goodness of fit between the *The Cannon*-predicted spectrum and the corresponding *Gaia*-RVS spectrum for each star. In this analysis, $v \sin i$ is included primarily as an additional label to improve the spectral modelling, and negative inferred values should therefore be interpreted as consistent with $v \sin i = 0$. For scientific analyses, we recommend selecting sources with `flag_cannon` $=$ `0`. For studies involving metallicity or individual elemental abundances, we additionally recommend applying the corresponding quality criterion (`flag_FeH` $=$ `0` and `flag_XFe` $=$ `0`, respectively). Further sample refinement may be performed by referring to Fig. C1, which shows the expected label uncertainties as a function of RVS spectral SNR.

Table A2 lists the schema of the catalogue containing chemical similarity-based GSE classifications and the corresponding orbital parameters. The column `Prob_GSE` gives the probability ($P_{\rm mem}$) that a star (satisfying `flag_cannon` $=$ `0`) is chemically similar to the GSE population using the MCMC model developed on *The Cannon*-predicted chemical abundances. The column `GSE_ref` identifies stars belonging to the cross-matched reference GSE sample (D. K. Feuillet et al. 2021). The kinematic quantities include the total orbital energy ($E$), angular momentum ($L_x$, $L_y$, $L_z$), the square root of the radial action ($\sqrt{J_R}$), Galactocentric velocities ($V_R$, $V_\phi$, $V_Z$), and orbital eccentricity ($e$), all derived from stellar orbital calculations using the MWPotential2014 Galactic potential.

**Table A1.** Table schema of the catalogue containing *The Cannon*-estimates.

| Column name | Unit | Description |
|---|---|---|
| `Gaia_dr3_id` | – | *Gaia* DR3 source ID |
| `RA` | deg | Right Ascension |
| `e_RA` | mas | Error in Right Ascension |
| `DEC` | deg | Declination |
| `e_DEC` | mas | Error in Declination |
| `plx` | mas | Parallax |
| `e_plx` | mas | Uncertainty in Parallax |
| `rvs_snr` | $\rm pix^{-1}$ | *Gaia* DR3 RVS signal-to-noise ratio |
| `flag_cannon` | – | Quality flag from *The Cannon* (`0`/`1`) |
| `Teff` | K | *Cannon*-estimated effective temperature |
| `e_Teff` | K | Uncertainty in effective temperature |
| `logg` | dex | *Cannon*-estimated surface gravity |
| `e_logg` | dex | Uncertainty in surface gravity |
| `vsini` | $\rm km\,s^{-1}$ | *Cannon*-estimated rotational velocity |
| `e_vsini` | $\rm km\,s^{-1}$ | Uncertainty in rotational velocity |
| `FeH` | dex | *Cannon*-estimated [Fe/H] |
| `e_FeH` | dex | Uncertainty in [Fe/H] |
| `flag_FeH` | – | *Cannon*-Flag for [Fe/H] (`0`/`1`) |
| `TiFe` | dex | *Cannon*-estimated [Ti/Fe] |
| `e_TiFe` | dex | Uncertainty in [Ti/Fe] |
| `flag_TiFe` | – | *Cannon*-Flag for [Ti/Fe] (`0`/`1`) |
| `SiFe` | dex | *Cannon*-estimated [Si/Fe] |
| `e_SiFe` | dex | Uncertainty in [Si/Fe] |
| `flag_SiFe` | – | *Cannon*-Flag for [Si/Fe] (`0`/`1`) |
| `CaFe` | dex | *Cannon*-estimated [Ca/Fe] |
| `e_CaFe` | dex | Uncertainty in [Ca/Fe] |
| `flag_CaFe` | – | *Cannon*-Flag for [Ca/Fe] (`0`/`1`) |
| `NiFe` | dex | *Cannon*-estimated [Ni/Fe] |
| `e_NiFe` | dex | Uncertainty in [Ni/Fe] |
| `flag_NiFe` | – | *Cannon*-Flag for [Ni/Fe] (`0`/`1`) |
| `ZrFe` | dex | *Cannon*-estimated [Zr/Fe] |
| `e_ZrFe` | dex | Uncertainty in [Zr/Fe] |
| `flag_ZrFe` | – | *Cannon*-Flag for [Zr/Fe] (`0`/`1`) |
| `CeFe` | dex | *Cannon*-estimated [Ce/Fe] |
| `e_CeFe` | dex | Uncertainty in [Ce/Fe] |
| `flag_CeFe` | – | *Cannon*-Flag for [Ce/Fe] (`0`/`1`) |
| `NdFe` | dex | *Cannon*-estimated [Nd/Fe] |
| `e_NdFe` | dex | Uncertainty in [Nd/Fe] |
| `flag_NdFe` | – | *Cannon*-Flag for [Nd/Fe] (`0`/`1`) |
| `rchi_sq` | – | Reduced-$\chi^2$ statistic |

*Notes.*(i) The columns `Gaia_dr3_id`, `RA`, `e_RA`, `DEC`, `e_DEC`, `plx`, `e_plx`, and `rvs_snr` are adopted from Gaia Collaboration (2023a).
(ii) A flag value of `0` indicates that the star lies within the corresponding stellar label manifold spanned by the training sample.

**Table A2.** Table schema of chemical similarity-based GSE membership probabilities and corresponding orbital parameters for stars in the catalogue.

| Column name | Unit | Description |
|---|---|---|
| `Gaia_dr3_id` | – | *Gaia* DR3 source ID |
| `Prob_GSE` | – | GSE Membership Probability ($P_{\rm mem}$) |
| `GSE_ref` | – | GSE Reference sample (`1` $=$ `True`) |
| `E` | $\rm km^2\,s^{-2}$ | Total orbital energy ($E$) |
| `Lx` | $\rm kpc\,km\,s^{-1}$ | Angular momentum ($L_x$) |
| `Ly` | $\rm kpc\,km\,s^{-1}$ | Angular momentum ($L_y$) |
| `Lz` | $\rm kpc\,km\,s^{-1}$ | Angular momentum ($L_z$) |
| `sqrt_Jr` | $\rm (kpc\,km\,s^{-1})^{1/2}$ | Square root of radial action ($\sqrt{J_r}$) |
| `Vr` | $\rm km\,s^{-1}$ | Radial velocity ($V_R$) |
| `Vphi` | $\rm km\,s^{-1}$ | Azimuthal velocity ($V_\phi$) |
| `Vz` | $\rm km\,s^{-1}$ | Vertical velocity ($V_z$) |
| `e` | – | Orbital eccentricity ($e$) |

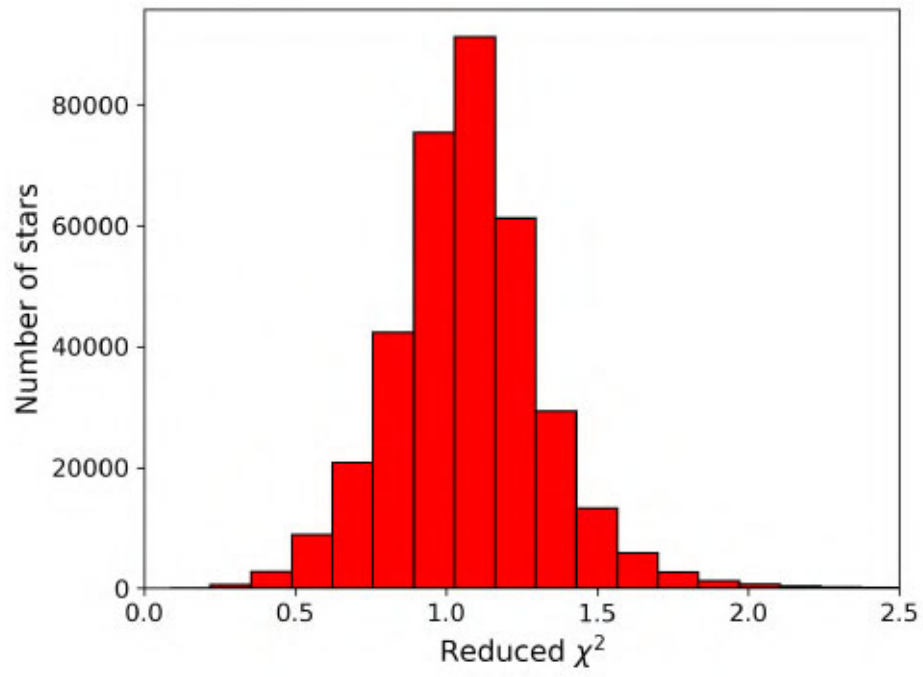

**Figure A1.** Distributions of reduced-$\chi^2$ values (`rchi_sq`) between the predicted spectra from *The Cannon* and the corresponding observed *Gaia*-RVS spectra for all the stars in our catalogue.

Fig. A1 shows the distribution of reduced `rchi_sq` values across the full sample of *Gaia*-RVS spectra. The reduced-$\chi^2$ statistic quantifies the goodness of fit between observed spectra and model spectra predicted by *The Cannon*; values close to unity indicate that the residual differences are broadly consistent with the estimated flux uncertainties. The distribution demonstrates that the majority of stars are fitted well by the model, indicating that *The Cannon* provides a robust representation of the observed RVS spectra across the parameter space explored in this work.

## APPENDIX B: The Cannon MODEL COEFFICIENTS

Fig. B1 shows the linear model coefficients for the stellar labels across the *Gaia* RVS spectral range, as inferred from modelling the training sample with *The Cannon*. These coefficients quantify the first-order sensitivity of the flux to label variations at each wavelength pixel.

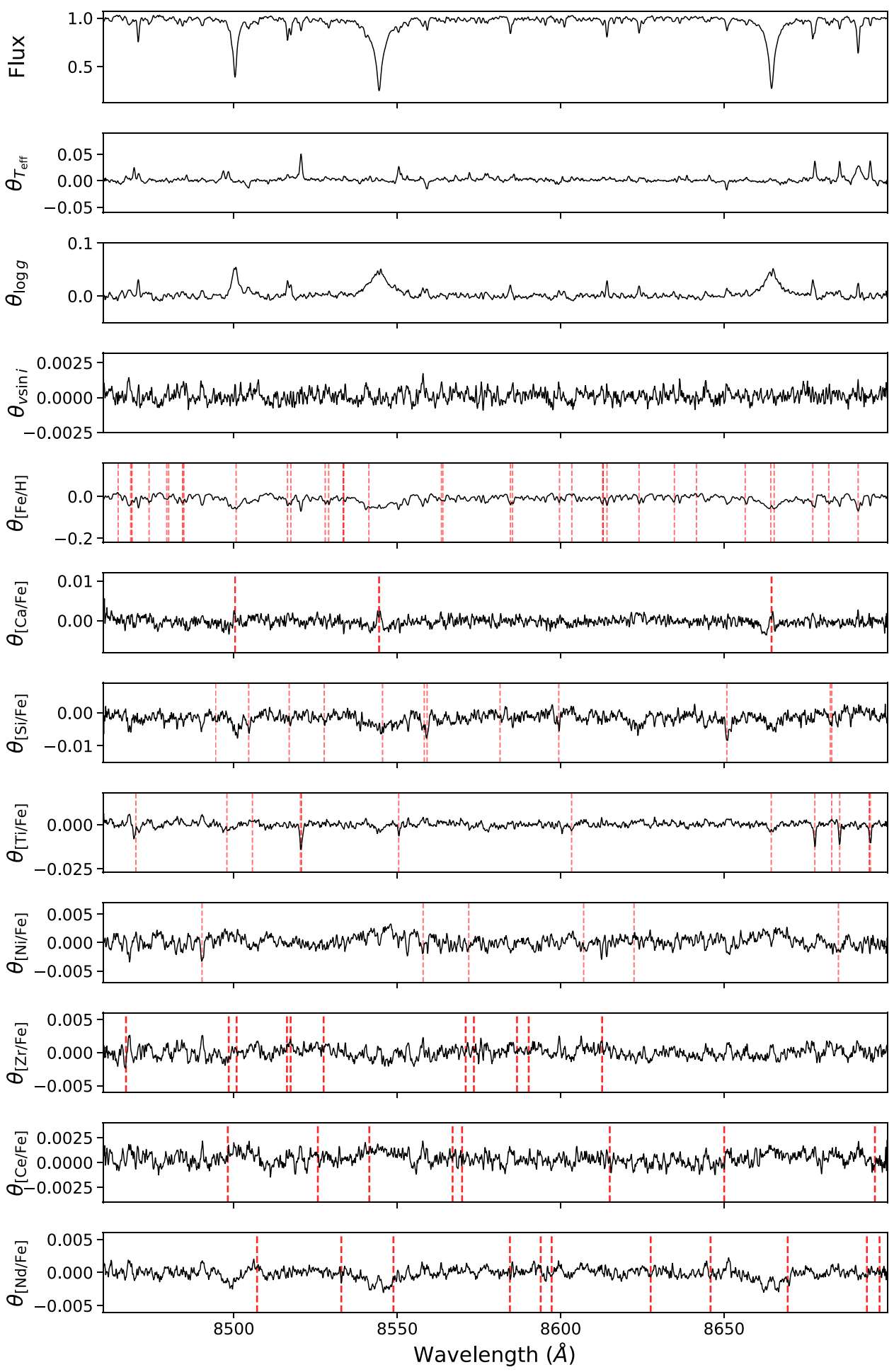

**Figure B1.** The topmost panel shows the *Gaia* RVS spectrum of a randomly selected star. The lower panels show the linear model coefficients from *The Cannon* ($\theta$) for the entire training sample of 2747 stars across the *Gaia* RVS spectral wavelength range for each stellar label. Red dashed vertical lines show the spectral lines associated with the corresponding abundance label.

## APPENDIX C: ERROR BUDGET IN STELLAR LABELS ACROSS Gaia-RVS SPECTRAL SNR

To quantify the dependence of the reported label uncertainties on spectral quality, we examined the total label uncertainties as a function of the *Gaia* RVS spectral SNR (SNR) for stars in the catalogue. Fig. C1 presents this relation for each stellar parameter and elemental abundance. In each panel, the two-dimensional density of stars is shown, and an exponential model of the form $\sigma_{\rm label} = Ae^{-(B*{\rm SNR})} + C$ is fitted to the data. Here, $A$ characterizes the amplitude of the increase in uncertainty toward lower SNRs, $B$ determines the rate at which the uncertainties improve with increasing SNR, and $C$ represents the precision floor reached at high SNR. The fits were obtained using a two-stage procedure in which median uncertainties were first computed in SNR bins and subsequently fitted with the exponential model.

Across all stellar parameters and elemental abundances, the uncertainties exhibit a systematic decline with increasing spectral SNR and converge toward a well-defined high-SNR precision limit. These empirical relations provide a quantitative framework for estimating the expected label precision as a function of spectral SNR and can be used to guide sample selection for applications requiring specific uncertainty thresholds.

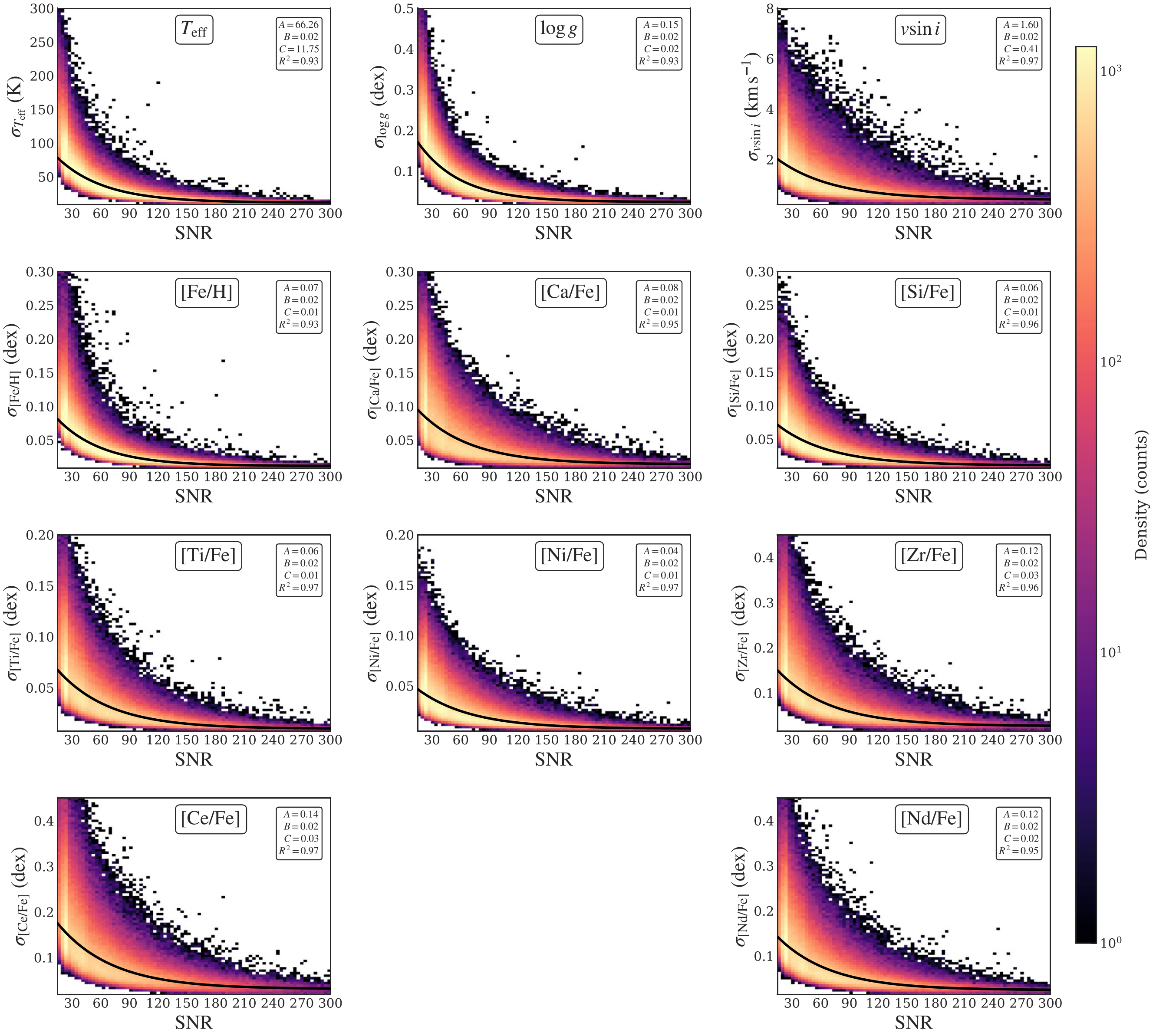

**Figure C1.** Uncertainties in the *The Cannon*-derived stellar labels as a function of *Gaia*-RVS spectral SNR for the stars in the catalogue. The stars in each panel satisfy the corresponding quality-flag criterion. The colour scale indicates the two-dimensional density of stars in each panel. The black curve shows an exponential fit of the form $\sigma_{\rm label} = Ae^{-(B*{\rm SNR})} + C$, where $\sigma$ denotes the reported uncertainty for each label. The fitted parameters and the coefficient of determination ($R^2$) are listed in each panel. In all cases, the uncertainties decrease with increasing spectral SNR and asymptotically approach a precision floor at high SNR.

## APPENDIX D: LINE PARAMETERS FOR CE, ZR, AND ND

Table D1 lists the vacuum and air wavelengths, species, excitation potentials (E.P.), and oscillator strengths ($\log gf$) for the Ce, Zr, and Nd from the Atomic and Molecular Line List Generator LINEMAKE. These sets of lines are contained within the *Gaia*-RVS wavelength range and were used in the *pseudo*-differential abundance analysis illustrated in Fig. 4. The 3 species (Ce, Zr, and Nd), follow the standard notation where neutral species are denoted as X.0 and ionized species as X.1, with X being the atomic number.

**Table D1.** Wavelengths, species, oscillator strengths, and excitation potentials for selected lines of Ce, Zr, and Nd. This linelist was created using LINEMAKE.

| λ vac. (Å) | λ air (Å) | Species | E.P. (eV) | log($gf$) (dex) |
|---|---|---|---|---|
| | | **Zr** | | |
| 8467.013 | 8464.687 | 40.0 | 0.650 | −2.08 |
| 8498.451 | 8496.117 | 40.0 | 2.124 | −1.37 |
| 8500.857 | 8498.522 | 40.0 | 0.622 | −2.22 |
| 8516.255 | 8513.916 | 40.0 | 0.632 | −3.11 |
| 8517.387 | 8515.047 | 40.0 | 2.114 | −1.64 |
| 8527.475 | 8525.133 | 40.1 | 2.408 | −1.00 |
| 8570.941 | 8568.587 | 40.0 | 0.730 | −2.80 |
| 8573.440 | 8571.085 | 40.0 | 1.529 | −2.07 |
| 8586.639 | 8584.281 | 40.0 | 1.873 | −1.32 |
| 8590.224 | 8587.865 | 40.0 | 1.481 | −2.12 |
| 8612.678 | 8610.313 | 40.0 | 1.481 | −2.27 |
| | | **Ce** | | |
| 8498.159 | 8495.825 | 58.0 | 0.294 | −1.65 |
| 8525.704 | 8523.362 | 58.0 | 0.294 | −1.74 |
| 8541.447 | 8539.101 | 58.1 | 0.175 | −3.20 |
| 8566.934 | 8564.581 | 58.0 | 0.274 | −1.80 |
| 8569.833 | 8567.479 | 58.0 | 0.493 | −1.51 |
| 8615.016 | 8612.650 | 58.0 | 0.466 | −1.39 |
| 8650.030 | 8647.655 | 58.0 | 0.552 | −1.30 |
| 8696.144 | 8693.756 | 58.1 | 2.279 | −1.26 |
| | | **Nd** | | |
| 8507.107 | 8504.770 | 60.1 | 1.356 | −1.68 |
| 8532.889 | 8530.545 | 60.1 | 0.064 | −3.61 |
| 8548.828 | 8546.480 | 60.1 | 2.019 | −1.23 |
| 8584.472 | 8582.114 | 60.1 | 1.698 | −1.46 |
| 8593.890 | 8591.530 | 60.1 | 1.357 | −1.65 |
| 8597.244 | 8594.883 | 60.1 | 1.140 | −1.86 |
| 8627.539 | 8625.170 | 60.1 | 1.877 | −1.45 |
| 8645.854 | 8643.480 | 60.1 | 1.199 | −1.29 |
| 8669.456 | 8667.075 | 60.1 | 1.263 | −1.89 |
| 8693.690 | 8691.303 | 60.1 | 1.349 | −1.73 |
| 8697.558 | 8695.170 | 60.1 | 1.698 | −1.38 |

## APPENDIX E: GSE-MEMBERSHIP PROBABILITY THRESHOLD SELECTION

Table E1 summarizes the classification metrics for different membership probability thresholds applied to the GSE candidate stars. Thresholds ranging from 50 per cent to 95 per cent were evaluated using the validation sample of 86 GSE stars (common to our catalogue and D. K. Feuillet et al. 2021), which was used to train the MCMC-based classifier. For each threshold, we tracked the true positives (TP), false positives (FP), and false negatives (FN). As the threshold increases, precision improves (fewer false positives), recall decreases (more missed GSE stars), and the total numbers of predicted and estimated true GSE stars decline, producing a progressively more conservative sample. We adopt a threshold of 55 per cent (highlighted in bold) for our analysis, as this value provides a balance between minimizing FP and retaining the majority of true GSE members, yielding a representative and reliable sample for subsequent population analysis.

**Table E1.** Membership classification metrics for different probability thresholds applied to GSE candidate stars. The threshold of 55 per cent (highlighted in bold) is adopted in our analysis, where the number of false positives is controlled while retaining the majority of true GSE members.

| Threshold (per cent) | TP | FP | FN | Precision (per cent) | Recall (per cent) | Predicted GSEs | Estimated True GSEs |
|---|---|---|---|---|---|---|---|
| 50 | 53 | 27 | 33 | 66.25 | 61.63 | 3595 | 2382 |
| **55** | **50** | **22** | **36** | **69.44** | **58.14** | **3295** | **2289** |
| 60 | 45 | 20 | 41 | 69.23 | 52.33 | 3042 | 2106 |
| 65 | 44 | 18 | 42 | 70.97 | 51.16 | 2769 | 1966 |
| 70 | 37 | 14 | 49 | 72.55 | 43.02 | 2481 | 1800 |
| 75 | 30 | 12 | 56 | 71.43 | 34.88 | 2235 | 1597 |
| 80 | 27 | 9 | 59 | 75.00 | 31.40 | 1985 | 1489 |
| 85 | 25 | 5 | 61 | 83.33 | 29.07 | 1705 | 1421 |
| 90 | 18 | 4 | 68 | 81.82 | 20.93 | 1364 | 1117 |
| 95 | 12 | 3 | 74 | 80.00 | 13.95 | 914 | 732 |

*Notes.* (i) *Threshold* – adopted membership probability threshold. (ii) TP – number of TP correctly identified in the validation sample of 86 confirmed GSE stars. (iii) FP – number of FP, i.e. stars incorrectly classified as GSE members. (iv) FN – number of FN, i.e. GSE stars missed by the classification. (v) Precision – fraction of predicted GSE members that are truly GSE, calculated as TP/(TP + FP). (vi) Recall – fraction of true GSE stars correctly identified, calculated as TP/(TP + FN). (vii) Predicted GSEs – total number of probable GSE candidates satisfying $P_{\mathrm{mem}} >$ Threshold, excluding the stars in common with the GSE reference sample from D. K. Feuillet et al. (2021). (viii) Estimated True GSEs – estimated true number of GSE stars, obtained by multiplying the precision by the number of predicted GSE candidates.

## APPENDIX F: CHEMICAL PROFILE OF PROBABLE GSE CANDIDATE MEMBERS

Fig. F1 shows the distribution of all possible pairwise chemical abundance ratios constructed from the elements [Fe, Ca, Ti, Nd, Zr, Ce] for the 2289 high-confidence probable GSE candidates ($P_{\mathrm{mem}} > 55$ per cent). These distributions are compared against those of Milky Way halo and thick disc stellar populations, as well as a reference sample of 86 stars common to our catalogue and the GSE sample from D. K. Feuillet et al. (2021).

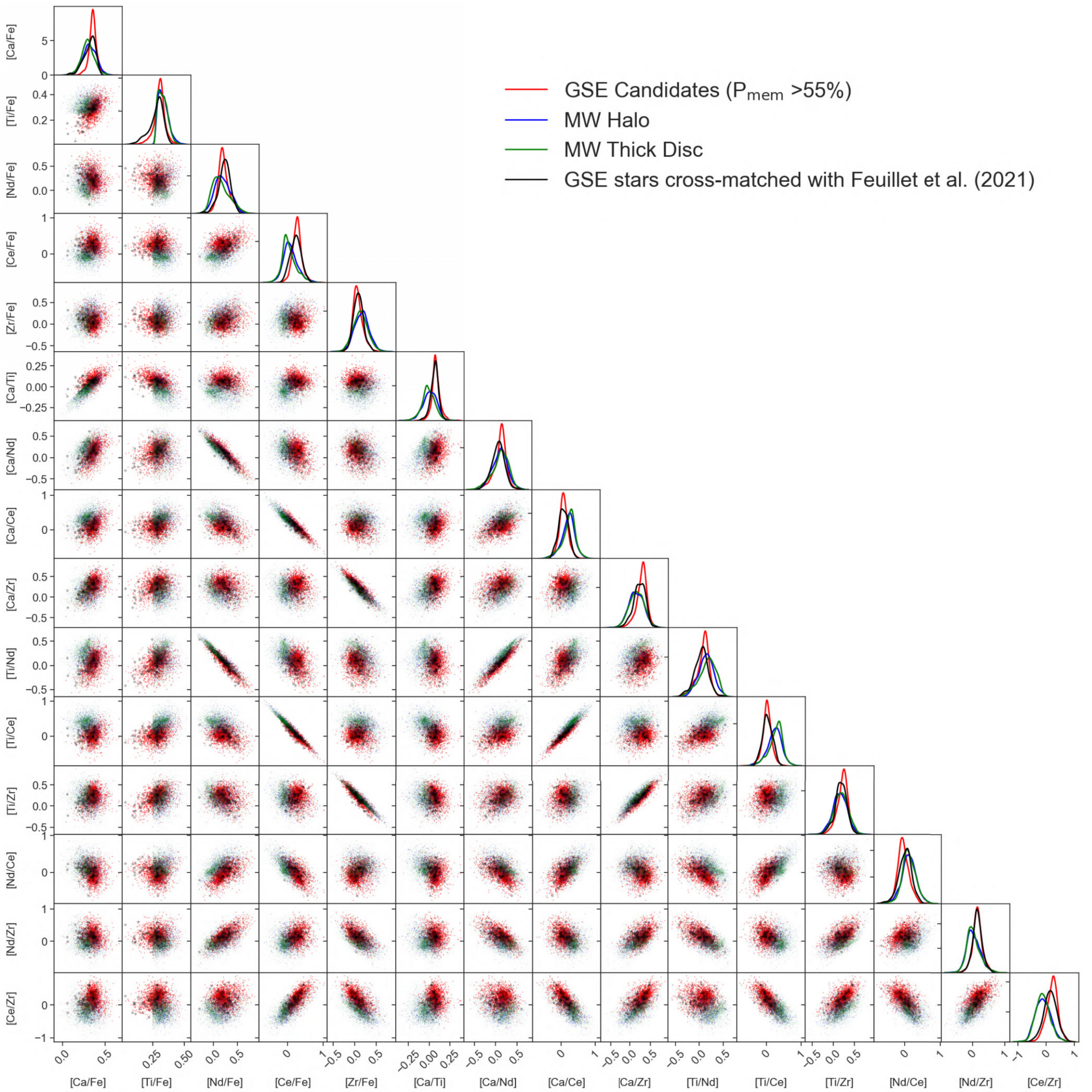

**Figure F1.** Distributions and pairwise correlations of elemental abundance ratios involving Fe, Ca, Ti, Nd, Ce, and Zr for four stellar populations: 2289 predicted GSE candidate stars with $P_{\rm mem} > 55$ per cent (red), kinematically selected halo stars (blue), thick disc stars (green), and a crossmatch of 86 GSE stars from D. K. Feuillet et al. (2021) with our catalogue of 314 010 *Gaia*-RVS stars (black) satisfying `flag_cannon = 0`. One-dimensional kernel density estimates are shown on the diagonal, and two-dimensional scatter plots appear below the diagonal. The distinct abundance trends, particularly in neutron-capture and $\alpha$-element ratios such as [Ca/Ti], [Ti/Ce], and [Nd/Zr], reveal clear chemical differentiation between the accreted GSE and *in situ* populations, highlighting their utility as tracers of stellar origin in the Milky Way.

[1] *School of Mathematical and Physical Sciences, Macquarie University, Sydney, NSW 2109, Australia*
[2] *Astrophysics and Space Technologies Research Centre, Macquarie University, Sydney, NSW 2109, Australia*
[3] *European Southern Observatory, Karl-Schwarzschild-Str 2, D-85748, Garching via Munich, Germany*
[4] *Lund Observatory, Division of Astrophysics, Department of Physics, Lund University, Box 118, SE-221 00, Lund, Sweden*
[5] *Research School of Astronomy and Astrophysics, The Australian National University, Canberra, ACT 2611, Australia*
[6] *ACCESS-NRI, Australian National University, Canberra, ACT 2601, Australia*
[7] *Observational Astrophysics, Department of Physics and Astronomy, Uppsala University, Box 516, SE-751 20, Uppsala, Sweden*
[8] *Lund Observatory, Department of Geology, Sölvegatan 12, SE-22362, Lund, Sweden*

[9]*Theoretical Astrophysics, Department of Physics and Astronomy, Uppsala University, Box 516, SE-751 20, Uppsala, Sweden*
[10]*School of Physics, University of New South Wales, Sydney, NSW 2052, Australia*
[11]*Faculty of Mathematics and Physics, University of Ljubljana, Jadranska 19, 1000, Ljubljana, Slovenia*
[12]*Sydney Institute for Astronomy, School of Physics, A28, The University of Sydney, Sydney, NSW 2006, Australia*
[13]*School of Physics and Astronomy, Monash University, Melbourne, VIC 3800, Australia*
[14]*Center for Computational Astrophysics, Flatiron Institute, 162 Fifth Avenue, New York, NY 10010, USA*
[15]*International Space Science Institute–Beijing, 1 Nanertiao, Zhongguancun, Hai Dian District, Beijing 100190, China*